\documentclass[twocolumn]{revtex4}
\usepackage[dvips]{graphicx}
\usepackage{amsfonts,amsmath}
\usepackage{float,color,array}

\begin{document}

\title{Spontaneous chirality selection and nonreciprocal spin wave in breathing-kagome antiferromagnets at zero field}

\author{Kazushi Aoyama$^1$ and Hikaru Kawamura$^2$}

\date{\today}

\affiliation{ $^1$Department of Earth and Space Science, Graduate School of Science, Osaka University, Osaka 560-0043, Japan \\
$^2$Molecular Photoscience Research Center, Kobe University, Kobe 657-8501, Japan}

\begin{abstract}
It has been known that the spin-wave dispersion, which is usually symmetric in the momentum space with respect to ${\bf q}=0$, can be asymmetric in the presence of the Dzyaloshinskii-Moriya (DM) interaction and an applied magnetic field. Here, we theoretically demonstrate that in $J_3$-dominant classical Heisenberg antiferromagnets on the breathing kagome lattice, the asymmetric spin-wave dispersion appears in a chiral phase due to non-uniform geometric phases acquired in the spin-wave propagation processes. This points to the emergence of a nonreciprocal spin wave in the absence of both the DM interaction and the magnetic field. Reflecting the asymmetry, positive-spin-chirality and negative-spin-chirality states, either one of which is selected in the low-temperature phase by the symmetry breaking, show different spin-wave dispersions, suggesting that the two energetically-degenerate chiral states can be distinguished by the spin-wave propagation.  
\end{abstract}

\maketitle
\section{Introduction}
The scalar spin chirality $\chi_{123}={\bf S}_1\cdot({\bf S}_2\times{\bf S}_3)$ defined by three localized spins ${\bf S}_1$, ${\bf S}_2$, ${\bf S}_3$ often plays an important role in magnetism. In particular, for $\chi_{ijk}$ defined on a triangle, when the total chirality summed over the whole system $\chi^{\rm T}=\sum_{i,j,k}\chi_{ijk}$ is nonzero, associated spin states can be distinguished by the sign of $\chi^{\rm T}$, analogous to Ising-type ferromagnetic states distinguished by the $\mathbb{Z}_2$ degrees of freedom, spin-up and spin-down. Since in metallic systems, $\chi^{\rm T}$ serves as an emergent fictitious magnetic field for conduction electrons, yielding the so-called topological Hall effect \cite{SkX_review_Nagaosa-Tokura_13, chiralityHall_Tatara_jpsj02, THE_Bruno_prl04, MnSi_Neubauer_09}, the sign difference in $\chi^{\rm T}$ is reflected in the electron propagation. Concerning the spin-wave propagation, it might also be affected by the spin chirality $\chi^{\rm T}$ \cite{SkX-Dyn_Mochizuki_prl12}, but the association between the spin wave and the sign of $\chi^{\rm T}$ is not yet well understood. In this paper, we theoretically investigate spin-wave dispersions in $\chi^{\rm T}>0$ and $\chi^{\rm T}<0$ phases realized as a result of spontaneous symmetry breaking at finite temperatures, picking up the zero-field $J_1$-$J_3$ Heisenberg model on the breathing kagome lattice \cite{KagomeSkX_AK_22, KagomeSkX_AK_23} as a platform for this research.    

A well-known example of the chiral phase with $\chi^{\rm T}\neq 0$ would be a skyrmion crystal (SkX) state. It is nowadays well-established that the SkX can be realized in the presence of the Dzyaloshinskii-Moriya (DM) interaction and an applied magnetic field  \cite{SkX_Bogdanov_89, SkX_Yi_09, SkX_Buhrandt_13, MnSi_Muhlbauer_09, FeCoSi_Yu_10, FeGe_Yu_11,  Cu2OSeO3_Seki_12, CoZnMn_Tokunaga_15, GaV4S8_Kezsmarki_15, GaV4Se8_Fujima_17, GaV4Se8_Bordacs_17, VOSe2O5_Kurumaji_17, AntiSkX_Nayak_17, EuPtSi_Kaneko_19}. Of recent particular interest would be SkX's of non-DM origins \cite{Gd2PdSi3_Kurumaji_19, GdRuAl_Hirschberger_natcom_19, GdRu2Si2_Khanh_20, EuAl4_Shang_prb_21,EuAl4_Takagi_natcom_22} such as magnetic frustration \cite{SkX_Okubo_12, SkX_Leonov_15, SkX-RKKY_Wang_20, SkX-RKKY_Mitsumoto_21, Kawamura_review} and a Fermi surface nesting \cite{SkX_top2_Ozawa_prl_17, SkX-RKKY_Hayami_17}. A crucial difference between the two is that in the DM systems, the sign of $\chi^{\rm T}$ is determined by the DM interaction from the beginning, whereas in the non-DM systems, the sign of $\chi^{\rm T}$ can be positive or negative as the Hamiltonian is invariant under the spin-space parity transformation by which $\chi^{\rm T}>0$ and $\chi^{\rm T}<0$ states are interchanged. In the latter case, either one of the two, $\chi^{\rm T}>0$ or $\chi^{\rm T}<0$, is selected by the spontaneous symmetry breaking. Previously, we theoretically showed that a breathing bond-alternation of the lattice can also induce the chiral order in the absence of the DM interaction; a miniature version of the SkX \cite{KagomeSkX_AK_22, KagomeSkX_AK_23} and a three-dimensional analogue of the SkX, a hedgehog lattice \cite{hedgehog_AK_prb_21, hedgehog_AK_prb_22}, can be realized on the breathing kagome and pyrochlore lattices, respectively. Since the miniature SkX state emerges even at zero field, influence of the spontaneous chirality selection could directly be discussed with the spin space being kept isotropic. In this work, we examine how the sign of the spin chirality $\chi^{\rm T}$ affects the spin-wave propagation in the zero-field Heisenberg antiferromagnet on the breathing kagome lattice.

\begin{figure*}[t]
\begin{center}
\includegraphics[scale=0.75]{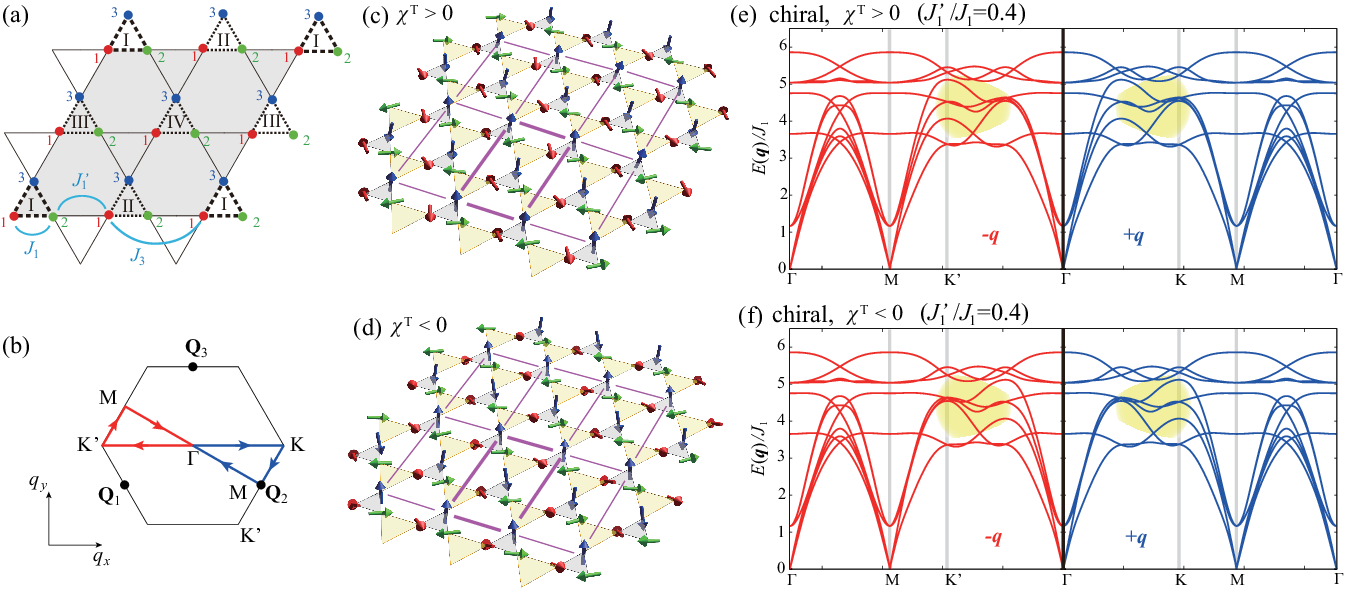}
\caption{(a) Breathing kagome lattice with antiferromagnetic interactions $J_1$, $J_1^\prime$, and $J_3$. A gray-colored region indicates the magnetic unit cell of the 12-sublattice triple-${\bf Q}$ state containing the four small triangles I, II, III, and IV. (b) The magnetic ordering vectors ${\bf Q}_1$, ${\bf Q}_2$, and ${\bf Q}_3$ and the Brillouin zone for the triple-${\bf Q}$ state with nonzero total chirality $\chi^{\rm T}\neq 0$. (c) and (d) Spin configurations in the positive chirality ($\chi^{\rm T}>0$) and negative chirality ($\chi^{\rm T}<0$) phases, respectively, where red, blue, and green arrows, respectively, represent spins on the corners 1, 2, and 3 of the small triangle shown in (a). A magenta rectangle indicates a miniature version of the (c) anti-skyrmion and (d) skyrmion. (e) and (f) Spin-wave dispersions $E({\bf q})$ in the $\chi^{\rm T}>0$ and $\chi^{\rm T}<0$ chiral phases for $J_1^\prime/J_1=0.4$, where the blue part on the right half  (red part on the left half) is $E({\bf q})$ along the path in the $q_x \geq 0$ ($q_x \leq 0$) region indicated by the blue (red) lines in (b). In each of (e) and (f), the asymmetric portions are highlighted in yellow. \label{fig1} }
\end{center}
\end{figure*}

The spin wave or the magnon is a collective magnetic excitation, and its dispersion $E({\bf q})$ is usually symmetric with respect to ${\bf q}=0$ (the $\Gamma$ point) in the momentum space, i.e., $E({\bf q})=E(-{\bf q})$. It is known, however, that in the presence of both the DM interaction and the magnetic field, the spin-wave dispersion in the field-induced ferromagnetic state becomes asymmetric, i.e., $E({\bf q}) \neq E(-{\bf q})$, for the ${\bf q}$ direction parallel to the field \cite{AsymDM_Kataoka_jpsj87} and such an asymmetry can be observed as a nonreciprocal spin-wave propagation \cite{AsymDM_Onose_prb15, AsymDM_Seki_prb16, AsymDM_Weber_prb18, AsymDM_Seki_natcom20}. In this case, the origin of the asymmetry can intuitively be understood as follows \cite{AsymDM_Seki_prb16}: the spin waves propagating in the ${\bf q}$ and $-{\bf q}$ directions along the field direction are characterized by the positive and negative chiralities, so that these two becomes nonequivalent under the chiral environment with the DM interaction. Then, the question is whether or not the asymmetry emerges as a result of chirality selection caused by spontaneous symmetry breaking in a situation where the Hamiltonian itself does not involve a chiral term or the DM interaction as well as the magnetic field.

By performing the spin-wave expansion and spin-dynamics simulations, we will show that in the chiral phase emerging in the $J_3$-dominant Heisenberg antiferromagnets on the breathing kagome lattice, the spin-wave dispersion becomes asymmetric in the absence of the DM interaction and the magnetic field, and resultantly, as shown in Figs. \ref{fig1} (e) and (f), the dispersions in the positive-spin-chirality ($\chi^{\rm T}>0$) and negative-spin-chirality ($\chi^{\rm T}<0$) states are different from each other. The asymmetry or nonreciprocity microscopically originates from geometric phases acquired in the spin-wave propagation processes. 

The outline of this paper is as follows: in Sec. II, we introduce the model and perform the spin-wave expansion to investigate the magnetic excitations. This is followed by Sec. III in which we show that the spin-wave dispersion becomes asymmetric in the chiral phase on the breathing kagome lattice. The origin of the asymmetry is discussed in Sec. IV. We end the paper with summary and discussion in Sec. V. Details of the spin-wave expansion and the calculation method of the dynamical spin structure factor are provided in Appendices A and B, respectively.

\section{Theoretical framework}
\subsection{Model}
We start from the following spin Hamiltonian on the breathing kagome lattice consisting of corner-sharing small and large triangles [see Fig. \ref{fig1} (a)]: 
\begin{equation}\label{eq:Hamiltonian}
{\cal H}  = J_1 \sum_{\langle i,j \rangle_S} {\bf S}_i\cdot{\bf S}_j + J_1' \sum_{\langle i,j \rangle_L} {\bf S}_i\cdot{\bf S}_j + J_3\sum_{\langle \langle  i,j \rangle \rangle } {\bf S}_i\cdot{\bf S}_j ,
\end{equation}
where ${\bf S}_i$ is a classical Heisenberg spin, $J_1>0$ and $J_1^\prime>0$ are the nearest neighbor (NN) antiferromagnetic interactions on the small and large triangles, respectively, and $J_3>0$ is the third NN antiferromagnetic interaction along the bond direction. In the model, the effect of the breathing lattice structure is taken into account in the form of the nonequivalent $J_1$ and $J_1^\prime$, and the real-space length difference is ignored for simplicity. 

For relatively strong $J_3$, a 12-sublattice triple-${\bf Q}$ state characterized by the three ordering vectors ${\bf Q}_1=\frac{\pi}{2a}(-1,-\frac{1}{\sqrt{3}})$, ${\bf Q}_2=\frac{\pi}{2a}(1,-\frac{1}{\sqrt{3}})$, and ${\bf Q}_3=\frac{\pi}{2a}(0,\frac{2}{\sqrt{3}})$ with side length of each triangle $a=1$ is favored. Although it takes a noncoplanar, coplanar, or collinear structure depending on the breathing parameter $J_1'/J_1$ and temperature $T$ \cite{KagomeSkX_AK_22}, the lowest temperature phase for $J_1^\prime/J_1 \neq 1$ has the noncoplanar structure whose energetically-degenerate two types of spin configurations are shown in Fig. \ref{fig1} (c) and (d). They correspond to the positive ($\chi^{\rm T} >0$) and negative ($\chi^{\rm T}<0$) chiral states, as inferred from the fact that they can be viewed as periodic arrays of the miniature anti-skyrmions and skyrmions [see the spin textures within the magenta rectangles in Figs. \ref{fig1} (c) and (d), respectively]. 

In the chiral phase, spins belonging to each of three corners of small triangles [see the same-color points or arrows in Figs. \ref{fig1} (a), (c), and (d)] constitute basically $\uparrow \downarrow \uparrow \downarrow$ chains running along the bond directions and the three chains [see the different color arrows in Figs. \ref{fig1} (c), and (d)] are superposed almost orthogonally. In the coplanar and collinear states, only the superposition angles among the three chains are different with each $\uparrow \downarrow \uparrow \downarrow$ structure remaining unchanged \cite{KagomeSkX_AK_22, KagomeSkX_AK_23}.

In a large parameter space available for the triple-${\bf Q}$ chiral state \cite{KagomeSkX_AK_22}, we use $J_3/J_1=1.2$ throughout this paper. We note that such a $J_3$-dominant situation seems to be realized in the {\it uniform}  kagome antiferromagnet BaCu$_3$V$_2$O$_8$(OD)$_2$ \cite{CoplanarOct_Boldrin_prl_18}. 
Below, we will discuss the dispersion of the magnetic excitations in the chiral state and for reference, in the coplanar and collinear states as well. 

\subsection{Spin-wave expansion}
The information of the magnetic excitations can be obtained by solving the spin-dynamics equation
\begin{equation}\label{eq:Bloch}
\frac{d {\bf S}_i}{dt}  =  {\bf S}_i \times {\bf H}_i^{\rm eff}, \quad {\bf H}_i^{\rm eff} = -\sum_j J_{ij} {\bf S}_j,
\end{equation}
where $J_{ij}$ takes $J_1$, $J_1^\prime$, and $J_3$ for the corresponding site pairs $(i,j)$. We perform the spin-wave expansion ${\bf S}_i=\overline{\bf S}_i + \delta {\bf S}_i$ with the ground-state spin configuration $\overline{\bf S}_i$ which, for the $\chi^{\rm T}>0$ state with the anti-SkX structure shown in Fig. \ref{fig1} (c), is given by (see Supplemental Material in Ref. \cite{KagomeSkX_AK_22})
\begin{eqnarray}\label{eq:spin_conf}
&&\mspace{-25mu} \left\{ \begin{array}{ll}
\overline{\bf S}_{{\rm I}1} = c\hat{P}_1+s (\hat{P}_2+\hat{P}_3), & \overline{\bf S}_{{\rm II}1} =  -c\hat{P}_1-s (\hat{P}_2-\hat{P}_3), \\
\overline{\bf S}_{{\rm III}1} = -c\hat{P}_1+s (\hat{P}_2-\hat{P}_3), & \overline{\bf S}_{{\rm IV}1} =  c\hat{P}_1-s (\hat{P}_2+\hat{P}_3),
\end{array} \right . \nonumber\\
&&\mspace{-25mu} \left\{ \begin{array}{cc}
\overline{\bf S}_{{\rm I}2} = -c\hat{P}_2-s (\hat{P}_1-\hat{P}_3), & \overline{\bf S}_{{\rm II}2} = c\hat{P}_2+s (\hat{P}_1+\hat{P}_3), \\
\overline{\bf S}_{{\rm III}2} = -c\hat{P}_2+s (\hat{P}_1-\hat{P}_3), & \overline{\bf S}_{{\rm IV}2} = c\hat{P}_2-s (\hat{P}_1+\hat{P}_3),
\end{array} \right . \nonumber\\
&&\mspace{-25mu} \left\{ \begin{array}{cc}
\overline{\bf S}_{{\rm I}3} =  -c\hat{P}_3-s (\hat{P}_1-\hat{P}_2), & \overline{\bf S}_{{\rm II}3} = -c\hat{P}_3+s (\hat{P}_1-\hat{P}_2), \nonumber\\
\overline{\bf S}_{{\rm III}3} =  c\hat{P}_3+s (\hat{P}_1+\hat{P}_2), & \overline{\bf S}_{{\rm IV}3} = c\hat{P}_3-s (\hat{P}_1+\hat{P}_2)
\end{array} \right . \\
\end{eqnarray}   
with $c=\frac{1}{\sqrt{1+2\varepsilon^2}}$, $s=\frac{\varepsilon}{\sqrt{1+2\varepsilon^2}}$,
\begin{eqnarray}
\varepsilon &=& \frac{ J_1-J_1'}{-\lambda+J_1+J_1'}, \nonumber\\
\lambda &=& \frac{J_1+J_1'-4J_3-\sqrt{(J_1+J_1'+4J_3)^2+8(J_1-J_1')^2}}{2} \nonumber
\end{eqnarray}
and mutually orthogonal three polarization vectors $\hat{P}_1$, $\hat{P}_2$, and $\hat{P}_3$. Here, $\overline{\bf S}_{\mu l}$ ($\mu =$I, II, III, and IV, and $l=$1, 2, and 3) denotes the spin at the corner $l$ of the small triangle $\mu$ in the magnetic unit cell [see Fig. \ref{fig1} (a)]. The spin configuration for the $\chi^{\rm T}<0$ state with the SkX structure shown in Fig. \ref{fig1} (d) is obtained by merely replacing $\hat{P}_3$ with $-\hat{P}_3$ in Eq. (\ref{eq:spin_conf}). 

Noting that $\varepsilon \neq 0$ and $s \neq 0$ only in the breathing case of $J_1^\prime/J_1\neq 1$, we find that the same-corner spins $\overline{\bf S}_{{\rm I}l}$, $\overline{\bf S}_{{\rm II}l}$, $\overline{\bf S}_{{\rm III}l}$, and $\overline{\bf S}_{{\rm IV}l}$ in Eq. (\ref{eq:spin_conf})  [the same color spins in Figs. \ref{fig1} (c) and (d)] are slightly tilted from the main polarization axis $\hat{P}_l$ due to the $s \propto \varepsilon$ term. By contrast, in the uniform case of $J_1^\prime/J_1=1$, we have $\varepsilon=0$, so that these spins constitute the perfectly collinear $\uparrow \downarrow \uparrow \downarrow$ chains. This difference between the breathing and uniform cases will turn out to be important.

Now, we consider the excitation from the ground state. At each of the 12 sublattices, the quantization axis (the $\tilde{S}^z_{\mu l}$ direction) is taken parallel to $\overline{\bf S}_{\mu l}$, and the fluctuations $\delta \tilde{S}^{\pm}_{\mu l}\equiv  \delta \tilde{S}^x_{\mu l} \pm i \delta \tilde {S}^y_{\mu l}$ are perpendicular to $\overline{\bf S}_{\mu l}$. As detailed in Appendix A, by introducing $\mbox{\boldmath $\Phi$}_{\bf q}= ^t( \mbox{\boldmath $\varphi$}^{+}_{{\rm I}, {\bf q}}, \mbox{\boldmath $\varphi$}^{+}_{{\rm II}, {\bf q}}, \mbox{\boldmath $\varphi$}^{+}_{{\rm III}, {\bf q}}, \mbox{\boldmath $\varphi$}^{+}_{{\rm IV}, {\bf q}}, \mbox{\boldmath $\varphi$}^{-}_{{\rm I}, {\bf q}}, \mbox{\boldmath $\varphi$}^{-}_{{\rm II}, {\bf q}}, \mbox{\boldmath $\varphi$}^{-}_{{\rm III}, {\bf q}}, \mbox{\boldmath $\varphi$}^{-}_{{\rm IV}, {\bf q}} )$ with $ \mbox{\boldmath $\varphi$}^{\pm}_{\mu, {\bf q}} =( \delta \tilde{S}^\pm_{\mu 1,{\bf q}}, \delta \tilde{S}^\pm_{\mu 2,{\bf q}}, \delta \tilde{S}^\pm_{\mu 3,{\bf q}} )$ consisting of the Fourier components of $\delta \tilde{S}^{\pm}_{\mu l}  = \sum_{\bf q} \delta \tilde{S}^{\pm}_{{\mu l},{\bf q}} e^{i {\bf q}\cdot{\bf r}_i}$, we can rewrite Eq. (\ref{eq:Bloch}) into the linearized 24$\times$24 matrix form as 
\begin{equation}\label{eq:Bloch_linear}
\mspace{-15mu}\frac{d}{dt} \mbox{\boldmath $\Phi$}_{\bf q}   = i H_{\bf q} \mbox{\boldmath $\Phi$}_{\bf q}, \quad H_{\bf q}= \left(\begin{array}{cc}
A_{\bf q} & B_{\bf q} \\
-B^\ast_{-{\bf q}} & -A^\ast_{-{\bf q}} 
\end{array} \right),   
\end{equation}
\begin{equation}\label{eq:M44}
A_{\bf q} = \left( \begin{array}{cccc}
D_1 & C_1 & C_3 & C_2 \\
C_1^\dagger & D_2 & C_2^\dagger & C_3 \\
C_3^\dagger & C_2 & D_3 & C_1^\dagger \\
C_2^\dagger & C_3^\dagger & C_1 & D_4 
\end{array}\right), \, B_{\bf q} = \left( \begin{array}{cccc}
Z_1 & V_1^- & V_3^- & V_2^+ \\
V_1^+ & Z_2 & V_2^- & V_3^- \\
V_3^+ & V_2^+ & Z_3 & V_1^+ \\
V_2^- & V_3^+ & V_1^- & Z_4 
\end{array}\right), 
\end{equation}
\begin{widetext}
\begin{eqnarray}
&& D_\mu = \left( \begin{array}{ccc}
a & s^\mu_1\gamma_+ G_{1{\bf q}} &  s^\mu_3\gamma_+^\ast G_{3{\bf q}}^\ast  \nonumber\\
s^\mu_1\gamma_+^\ast G_{1{\bf q}}^\ast & a & s^\mu_2 \gamma_+ G_{2{\bf q}} \nonumber\\
s^\mu_3 \gamma_+ G_{3{\bf q}} & s^\mu_2 \gamma_+^\ast G_{2{\bf q}}^\ast & a  \nonumber
\end{array}\right), 
C_1 = \left( \begin{array}{ccc}
-s^2 T_{1{\bf q}} & -\gamma_- G_{1{\bf q}}^{\prime \, \ast} &  0 \nonumber\\
\gamma_-^\ast G_{1{\bf q}}^\prime & s^2 T_{1{\bf q}} & 0 \nonumber\\
0 &0 & 0  \nonumber
\end{array}\right), 
C_2 = \left( \begin{array}{ccc}
0 & 0 & 0 \nonumber\\
0 & -s^2 T_{2{\bf q}} & \gamma_- G_{2{\bf q}}^{\prime \, \ast}  \nonumber\\
0 & -\gamma_-^\ast G_{2{\bf q}}^\prime & s^2 T_{2{\bf q}} \nonumber
\end{array}\right), \nonumber\\
&& C_3 = \left( \begin{array}{ccc}
s^2 T_{3{\bf q}} & 0 & -\gamma_-^\ast G_{3{\bf q}}^{\prime}  \nonumber\\
0 & 0 & 0 \nonumber\\
\gamma_- G_{3{\bf q}}^{\prime \, \ast} &0 & -s^2 T_{3{\bf q}}  \nonumber
\end{array}\right),  
Z_\mu = i b_+ \left( \begin{array}{ccc}
0 & s^\mu_1 G_{1{\bf q}} &  s^\mu_3 G_{3{\bf q}}^\ast  \nonumber\\
s^\mu_1 G_{1{\bf q}}^\ast & 0 & s^\mu_2 G_{2{\bf q}} \nonumber\\
s^\mu_3 G_{3{\bf q}} & s^\mu_2 G_{2{\bf q}}^\ast & 0  \nonumber
\end{array}\right), 
V_1^\pm =  \left( \begin{array}{ccc}
-\gamma_0 T_{1{\bf q}} & \pm i b_- G_{1{\bf q}}^{\prime \, \ast} &  0 \nonumber\\
\mp i b_- G_{1{\bf q}}^\prime & -\gamma_0^\ast T_{1{\bf q}} & 0 \nonumber\\
0 &0 & 0  \nonumber
\end{array}\right), \nonumber\\
&& V_2^\pm =  \left( \begin{array}{ccc}
0 & 0 & 0 \nonumber\\
0 & -\gamma_0 T_{2{\bf q}} & \pm i b_- G_{2{\bf q}}^{\prime \, \ast}  \nonumber\\
0 & \mp i b_- G_{2{\bf q}}^\prime & -\gamma_0^\ast T_{2{\bf q}} \nonumber
\end{array}\right), 
V_3^\pm =  \left( \begin{array}{ccc}
-\gamma_0^\ast T_{3{\bf q}} & 0 & \pm i b_- G_{3{\bf q}}^{\prime}  \nonumber\\
0 & 0 & 0 \nonumber\\
\mp i b_- G_{3{\bf q}}^{\prime \, \ast} &0 & -\gamma_0 T_{3{\bf q}}  \nonumber
\end{array}\right)  \nonumber
\end{eqnarray}
\end{widetext}
with
\begin{eqnarray}\label{eq:coefficient}
&& a = -2[\varepsilon(J_1-J_1^\prime)+2J_3], \quad  b_\pm = \frac{c^2+1}{4}\pm cs, \nonumber\\
&& \gamma_\pm =  \frac{c^2-1}{4} \pm cs + i \Big(\frac{c}{2}\mp s \Big) , \quad \gamma_0 = c+is^2 \nonumber\\
&& G^{(\prime)}_{i {\bf q}} = J_1^{(\prime)} e^{i{\bf q}\cdot {\bf e}_i}, \quad T_{i{\bf q}} = 2J_3\cos(2{\bf q}\cdot{\bf e}_i), 
\end{eqnarray}
and the sign factor $s^\mu_{\nu} = 1-2(\delta_{\mu\nu}+\delta_{\mu 4})=\pm 1$. Then, the problem is reduced to diagonalize the 24$\times$24 matrix; the normal-mode frequency $\omega$ defined by $\mbox{\boldmath $\Phi$}_{\bf q} = \sum_\omega \mbox{\boldmath $\Phi$}_{\bf q}(\omega) \, e^{-i\omega t}$ corresponds to the eigen value $E({\bf q})$ of the matrix $H_{\bf q}$.  

In Eq. (\ref{eq:Bloch_linear}), the ${\bf q}$ dependence of $H_{\bf q}$ comes from $G_{i {\bf q}}$, $G^\prime_{i {\bf q}}$, and $T_{i{\bf q}}$ involving the exchange interactions $J_1$, $J_1^\prime$, and $J_3$, respectively, where in Eq. (\ref{eq:coefficient}), ${\bf e}_1=(1,0)$, ${\bf e}_2=(-\frac{1}{2}, \frac{\sqrt{3}}{2})$, and ${\bf e}_3=(-\frac{1}{2},-\frac{\sqrt{3}}{2})$ are the real-space vectors along the bond directions (see Fig. \ref{fig4}).  
Since the matrix elements involving $G_{i {\bf q}}^{(\prime)}$ and $T_{i{\bf q}}$ describes the spin-wave propagation or the magnon hopping between site pairs connected by $J_1^{(\prime)}$ and $J_3$, respectively, their complex-value prefactors $\gamma_\pm$, $\gamma_0$, and $i b_\pm$ can be interpreted as geometric phases acquired when the spin fluctuation $\delta {\bf S}_{\mu l}\perp \overline{\bf S}_{\mu l}$ propagates on the curved surface spanned by the spatially varying quantization axes $\overline{\bf S}_{\mu l}$ \cite{Berry_Bruno_book_07, Berry_Dugaev_prb_05}. 

In the uniform limit of $J_1^\prime/J_1 \rightarrow 1$ where $\varepsilon \rightarrow 0$, $c = \frac{1}{\sqrt{1+2\varepsilon^2}}\rightarrow 1$, and $s=\frac{\varepsilon}{\sqrt{1+2\varepsilon^2}} \rightarrow 0$, the phase factors for the NN pairs take the same pure imaginary value, i.e., $\gamma_\pm \rightarrow i/2=e^{i\frac{\pi}{2}}/2$ and $i b_\pm \rightarrow i/2=e^{i\frac{\pi}{2}}/2$, so that the geometric phase between the NN pairs is $\frac{\pi}{2}$, which originates from the fact that the neighboring spins are perfectly orthogonal to each other [see Eq. (\ref{eq:spin_conf}) with $\varepsilon \rightarrow 0$ ($s \rightarrow 0$)]. In addition, the phase factor between the third-nearest pairs takes the real value of $\gamma_0 \rightarrow 1=e^{i 0}$, as the corresponding spins are perfectly collinear. Thus, the geometric phase takes nontrivial values only in the breathing case of $J_1^\prime/J_1 \neq 1$ ($\varepsilon \neq 0$), which suggests that novel phenomena may emerge in the breathing system.
  
 \section{Result on the spin wave dispersion}
\begin{figure}[t]
\begin{center}
\includegraphics[width=\columnwidth]{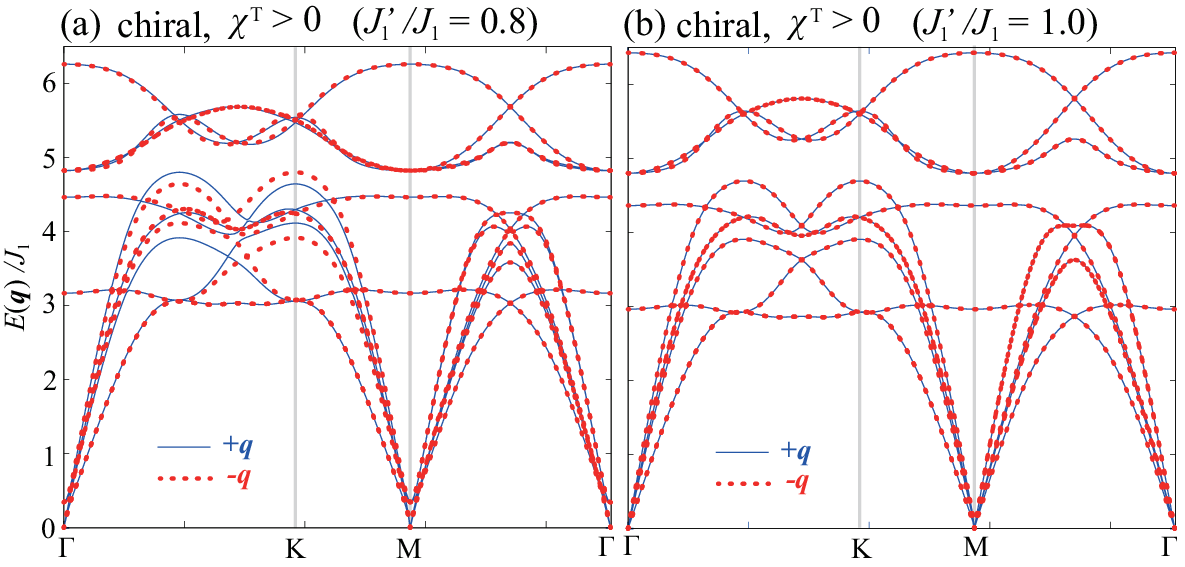}
\caption{Spin wave dispersion $E({\bf q})$ in the $\chi^{\rm T}>0$ phase for (a) $J_1^\prime/J_1=0.8$ and (b) $J_1^\prime/J_1=1$, where the results in the positive- and negative-$q_x$ regions [blue and red trajectories in Fig. \ref{fig1} (b)] are represented by blue and red-dashed curves, respectively, and the latter is folded back in the positive-$q_x$ region. \label{fig2} }
\end{center}
\end{figure}

Figures \ref{fig1} (e) and (f) show the spin-wave dispersion $E({\bf q})$ in the positive spin-chirality ($\chi^{\rm T}>0$) and negative spin-chirality ($\chi^{\rm T}<0$) states for $J_1^\prime/J_1=0.4$, respectively, where the blue part on the right half (red part on the left half) is $E({\bf q})$ obtained along the path of high-symmetry points in the $q_x \geq 0$ ($q_x \leq 0$) region of the Brillouin zone [see the blue (red) lines in Fig. \ref{fig1} (b)]. The difference between Figs. \ref{fig1} (e) and (f) can clearly be seen: the several spin-wave branches near the $K$ point are merged in the $\chi^{\rm T}>0$ phase, while not in the $\chi^{\rm T}<0$ phase. Thus, the chirality selection by the symmetry breaking affects the spin dynamics. Furthermore, by comparing the $+{\bf q}$ region to the $-{\bf q}$ region in each of Figs. \ref{fig1} (e) and (f) (see the yellow regions in each figure), we notice that the spin-wave dispersion is asymmetric with respect to the $\Gamma$ point. Since as inferred from the fact that the positive and negative chiral states are the time-reversal pair, $E(+{\bf q})$ in the $\chi^{\rm T}>0$ phase [the right blue part in Fig. \ref{fig1} (e)] corresponds to $E(-{\bf q})$ in the $\chi^{\rm T}<0$ phase [the left red part in Fig. \ref{fig1} (f)], we could conclude that the difference between the $\chi^{\rm T} >0$ and $\chi^{\rm T}<0$ cases originates from the asymmetric spin-wave dispersion in each chiral phase. When we take a closer look at the dispersion in Fig. \ref{fig1} (e), we also notice that the asymmetry disappears along the path connecting the $\Gamma$ and $M$ points, which can clearly be seen in Fig. \ref{fig2} (a). To understand why the dispersion becomes asymmetric, we examine the $J_1^\prime/J_1$ dependence of the dispersion $E({\bf q})$.
 
Figure \ref{fig2} shows the spin-wave dispersion $E({\bf q})$ in the $\chi^{\rm T}>0$ phase for (a) $J_1^\prime/J_1 = 0.8$ and (b) $J_1^\prime/J_1=1$, where the $-{\bf q}$ region indicated by red dotted curves is folded back to the $+{\bf q}$ region. One can see from Figs. \ref{fig1} (e), \ref{fig2} (a), and \ref{fig2} (b) that as the breathing structure becomes weaker, i.e., $J_1^\prime/J_1$ approaches 1, the asymmetry in the spin-wave dispersion becomes smaller and it vanishes in the uniform case of $J_1^\prime/J_1=1$. This suggests that the breathing bond alternation is essential for the asymmetric dispersion.   

For comparison, we also examine how the spin excitation looks in the collinear and coplanar states. Since these states are stable only at finite temperatures, we prepare finite-temperature equilibrium spin configurations in the Monte Carlo (MC) simulation and then, numerically integrate Eq. (\ref{eq:Bloch}) up to $t=t_0=1000/J_1$ with the MC snapshot as the initial state to obtain the dynamical spin structure factor $S({\bf q}, \omega)= |\frac{1}{t_0}\int dt \, \frac{1}{N}\sum_i {\bf S}_i e^{-i ({\bf q}\cdot {\bf r}_i- \omega t)}|^2$, where $N=3L^2$ is the total number of spins (for details, see Appendix B). $S({\bf q},\omega)$ is a physical quantity reflecting the dispersion of the magnetic excitations.

\begin{figure}[t]
\begin{center}
\includegraphics[width=\columnwidth]{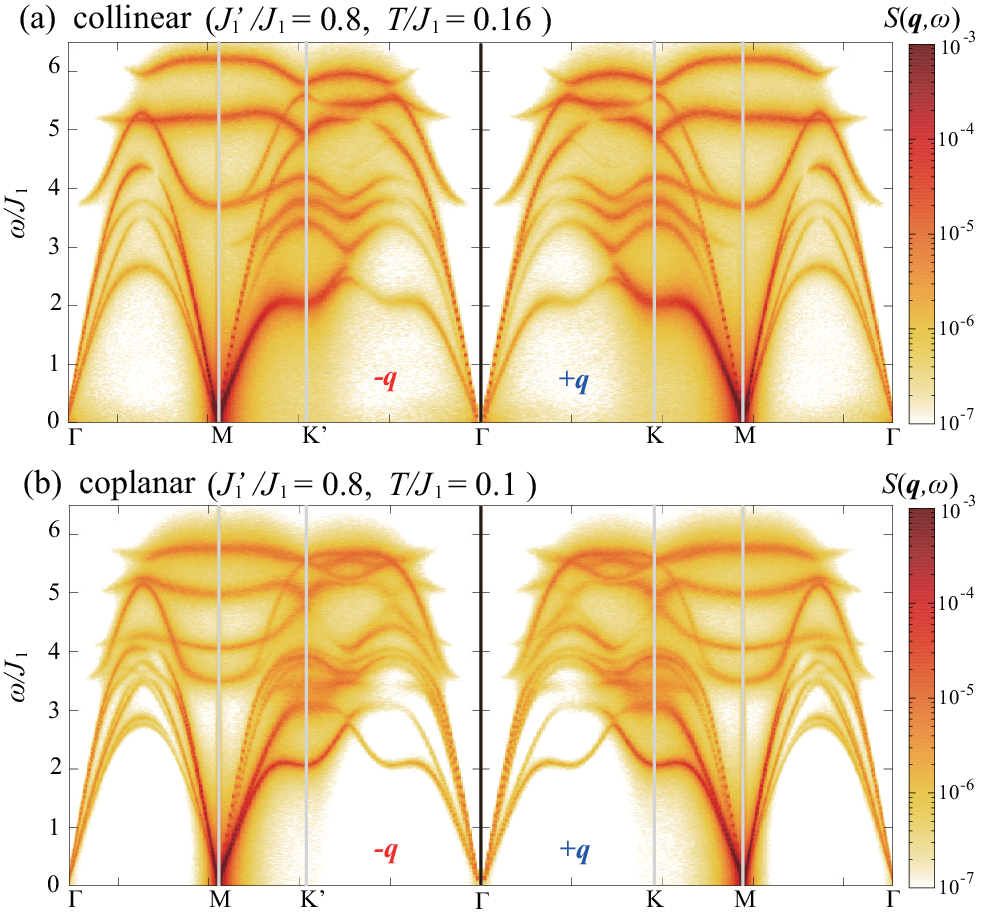}
\caption{Dynamical spin structure factor $S({\bf q}, \omega)$ in (a) the collinear state at $T/J_1=0.16$ and (b) the coplanar state at $T/J_1=0.1$ in the breathing case of $J_1^\prime/J_1=0.8$, where the results are obtained in the hybrid MC and spin-dynamics simulation for $L=288$. The results are averaged over four independent runs to reduce the thermal noise, and the high-intensity Bragg contribution just at ${\bf Q}_2$ is removed to focus on low-intensity fluctuations. \label{fig3} }
\end{center}
\end{figure}

\begin{figure}[t]
\begin{center}
\includegraphics[width=\columnwidth]{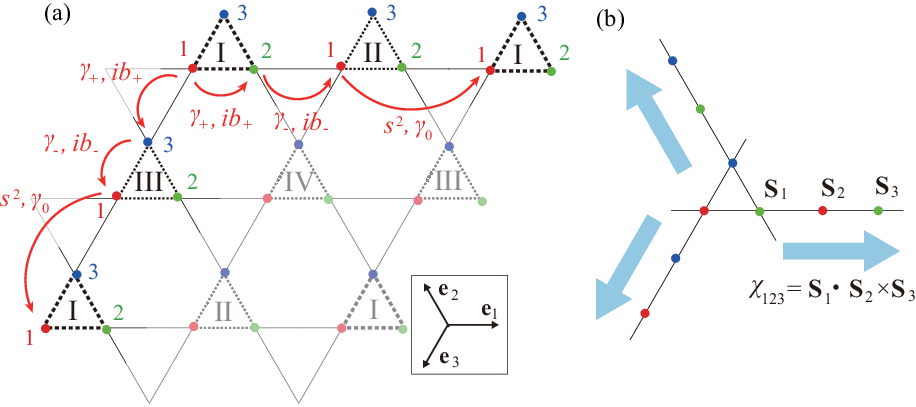}
\caption{(a) Propagation processes of the spin waves or the magnons relevant to the nonreciprocity. Each process indicated by a red arrow involves a geometric-phase factor such as $\gamma_\pm$ and $ib_\pm$. (b) Spin chirality defined in a straight line, where arrows indicate the bond directions ${\bf e}_1$, ${\bf e}_2$, and ${\bf e}_3$ in (a).  \label{fig4} }
\end{center}
\end{figure}

Figure \ref{fig3} shows $S({\bf q},\omega)$ obtained in the (a) collinear and (b) coplanar states in the breathing case of $J_1^\prime/J_1=0.8$.  
In both the collinear and coplanar states, $S({\bf q},\omega)$ is symmetric with respect to the $\Gamma$ point, which is in contrast to the situation in the chiral phase where the asymmetric structure can clearly be seen in the spin-wave dispersion $E({\bf q})$ [see Fig. \ref{fig2} (a)] and resultantly, in the dynamical spin structure factor $S({\bf q},\omega)$ as well (see Fig. \ref{fig:SM1} in Appendix B).
Since the asymmetric dispersion appears only in the chiral phase, both the emergent chiral order and the breathing lattice structure are important for the emergence of the asymmetric spin-wave dispersion pointing to the nonreciprocal spin-wave propagation.   

\section{Origin of the nonreciprocal spin wave}
To understand the microscopic origin of the asymmetry or the nonreciprocity, we consider the quantity $F_{\bf q}^{(n)}={\rm tr}\big[ H_{\bf q} ^n\big]-{\rm tr}\big[ H_{-{\bf q}} ^n \big]$ which takes a nonzero value when the spin-wave dispersion is asymmetric \cite{AsymMag_Hayami_prb21, AsymMag_Hayami_prb22}. Noting that Eq. (\ref{eq:Bloch_linear}) can formally be solved as $\mbox{\boldmath $\Phi$}_{\bf q}(t)=e^{iH_{\bf q}t}\mbox{\boldmath $\Phi$}_{\bf q}(0)$, we finds that $H_{\bf q} ^n$ is associated with $n$ times hoppings of the magnon and thus, $F_{\bf q}^{(n)}$ is an indicator of the nonreciprocity in the magnon propagation process. Since $H_{\bf q}$ is the 24$\times$24 matrix, $F_{\bf q}^{(n)}$ contains so many terms which might be canceled out, we perform the numerical evaluation of $F_{\bf q}^{(n)}$ instead of a direct analytical calculation, and then, try to extract important contributions. From the numerical calculation, it is found that a nonzero leading-order contribution is $F_{\bf q}^{(3)}$ and that when the matrix $H_{\bf q}$ is divided into 2$\times$2 blocks like Eq. (\ref{eq:Bloch_linear}), the dominant contribution in $F_{\bf q}^{(3)}$ comes from $-6{\rm Re}\Big( {\rm tr}\big[ A_{\bf q} B_{\bf q} B^\ast_{-{\bf q}} \big]- \big\{ {\bf q}\rightarrow -{\bf q}\big\}\Big)$. By further dividing the matrices $A_{\bf q}$ and $B_{\bf q}$ into 4$\times$4 blocks like Eq. (\ref{eq:M44}), i.e., dividing $[ A_{\bf q} B_{\bf q} B^\ast_{-{\bf q}}]$ into 4$\times$4 blocks, we find that the contributions from the first, second, third, and fourth blocks $[ A_{\bf q} B_{\bf q} B^\ast_{-{\bf q}}]_{11}$,  $[ A_{\bf q} B_{\bf q} B^\ast_{-{\bf q}}]_{22}$, $[ A_{\bf q} B_{\bf q} B^\ast_{-{\bf q}} ]_{33}$, and $[ A_{\bf q} B_{\bf q} B^\ast_{-{\bf q}} ]_{44}$ are equal to one another. Thus, as a representative example, we consider the first block component $-6{\rm Re}\Big( {\rm tr}\Big[ [A_{\bf q} B_{\bf q} B^\ast_{-{\bf q}}]_{11} \Big] \Big)$ which can analytically be calculated as  
\begin{eqnarray}\label{eq:nonreciprocity}
&& -6\times 8 {\rm Re}\big[ (b_+ \gamma_-  - b_- \gamma_+  )\gamma_0^\ast \big]  \sum_{i=1}^3 {\rm Im}\big[  T_{i{\bf q}} G_{i{\bf q}}G_{i{\bf q}}^\prime  \big] \nonumber\\
&& = 48 \,  J_1J_1^\prime J_3 \, s(c^4-s^4) \sum_{i=1}^3 \sin(4{\bf q}\cdot{\bf e}_i) .
 \end{eqnarray}
Since in Eqs. (\ref{eq:Bloch_linear}) and (\ref{eq:M44}), the first block matrix, e.g., $D_1$, acts on the first three components of $\mbox{\boldmath $\Phi$}_{\bf q}$, $\mbox{\boldmath $\varphi$}^+_{{\rm I},{\bf q}} = (\delta \tilde{S}_{{\rm I}1,{\bf q}}^+, \delta \tilde{S}_{{\rm I}2,{\bf q}}^+, \delta \tilde{S}_{{\rm I}3,{\bf q}}^+)$, Eq. (\ref{eq:nonreciprocity}) can be interpreted as the nonreciprocity of the spin-wave propagating from the sites I1, I2, and I3 to the neighboring same-sublattice sites by the hopping $J_1$, $J_1^\prime$, and $J_3$, each involving $\gamma_+$ or $ib_+$, $\gamma_-$ or $ib_-$, and $s^2$ or $\gamma_0$, respectively [see Fig. \ref{fig4} (a)]. In Eq. (\ref{eq:nonreciprocity}), the ${\bf q}$-dependent part becomes $\sum_{i=1}^3 \sin(4{\bf q}\cdot{\bf e}_i)=\sin(4q)-2\sin(2q)\neq 0$ and 0 for ${\bf q} =q (1,0) $ and $q(\frac{\sqrt{3}}{2},-\frac{1}{2})$, respectively, so that the nonreciprocity occurs in the $K$-point direction but not in the $M$-point direction, which is consistent with the results shown in Figs. \ref{fig1} (e) and \ref{fig2} (a). Furthermore, in the uniform case of $J_1^\prime /J_1 =1$ where $\varepsilon =0$ and resultantly $s=0$, the nonreciprocity (\ref{eq:nonreciprocity}) turns out to disappear as $b_+ \gamma_- - b_-\gamma_+ =s(-c+\frac{3c^2+1}{2}i)=0$. Thus, the nonequivalence of the factors $ib_\pm$ and $\gamma_\pm$, namely, the non-uniformity of the associated geometric phase in the hopping process, is essential for the asymmetric spin-wave dispersion. 

Now that the microscopic origin of the asymmetry is understood, we shall interpret it in terms of the real-space spin structure. First, from the viewpoint of symmetry, $E({\bf q})\neq E(-{\bf q})$ points to a real-space broken-inversion-symmetry, which may be captured by the spatial distribution of the scalar spin chirality $\chi_{ijk}$ \cite{AsymMag_Chernyshev_prb_16}. Actually in the present system, $\chi_{ijk}$'s on the small and large triangles are calculated from Eq. (\ref{eq:spin_conf}) as $(1-3\varepsilon^2-2\varepsilon^3)c^3$ and $(1-3\varepsilon^2+2\varepsilon^3)c^3$, respectively, so that the state is not invariant under the real-space inversion interchanging these chiralities. 
Although the nonreciprocal spin wave should have a root in this underlying chirality distribution with broken inversion symmetry, strictly speaking, it originates from the dynamical propagation process accompanied with the geometric phase, as discussed above. To take the propagation direction into account, as shown in Fig. \ref{fig4} (b), we consider $\chi_{ijk}$ defined in a straight line.     
In the chiral phase, the straight-line chirality can be calculated from Eq. (\ref{eq:spin_conf}) as $\chi_{ijk} = -2 \varepsilon (1-\varepsilon^2) \,c^3$ for any of the neighboring three spins aligned in the bond directions ${\bf e}_1$, ${\bf e}_2$, and ${\bf e}_3$. Thus, the spin waves propagating in the opposite directions feel the opposite spin-chiralities because of the trivial algebraic relation ${\bf S}_1 \cdot ({\bf S}_2 \times {\bf S}_3) =- {\bf S}_3 \cdot ({\bf S}_2 \times {\bf S}_1)$, which should result in the asymmetric spin-wave dispersion. By contrast, in the uniform case of $\varepsilon=0$, the straight-line spin-chirality is zero, so that there is no distinction between the propagations in the opposite directions. One can easily show that a similar argument can also be applied in the different system of an in-field easy-plane antiferromagnet \cite{AsymMag_Chernyshev_prb_16}, so that the straight-line spin-chirality might serve as a good indicator for the nonreciprocal spin waves in noncoplanar spin states.

\section{Summary and discussion}
In this work, we have theoretically shown that in the chiral phase emerging in the zero-field breathing-kagome antiferromagnets, the spin wave dispersion becomes asymmetric, and resultantly, the dispersions in the positive-spin-chirality ($\chi^{\rm T}>0$) phase with the miniature anti-SkX structure and in the negative-spin-chirality ($\chi^{\rm T}<0$) phase with the miniature SkX structure are different from each other. This suggests that the spin wave can distinguish the two energetically degenerate chiral states, similarly to the conduction electrons feeling the emergent fictitious magnetic field which can take a positive or negative value depending on the sign of $\chi^{\rm T}$. Similar asymmetric spin-wave dispersions have been discussed also in other DM-free systems \cite{AsymMag_Hayami_prb22, AsymMag_Chernyshev_prb_16, AsymMag_Chernyshev_prx19, AsymMag_Cheon_prb18} where spin-orbit couplings (SOCs) or a magnetic field plays an important role. In the present $J_3$-dominant breathing-kagome Heisenberg antiferromagnet without the SOC and the magnetic field, the dispersion becomes asymmetric due to both the emergent chiral order and the breathing bond alternation, which can be understood microscopically as a result of the non-uniform geometric phase acquired in the magnon hopping processes. The non-uniformity of the geometric phase is reflected in the spin chirality ${\bf S}_i\cdot({\bf S}_j \times {\bf S}_k)$ defined in a straight line. Whether or not such a description based on the geometric phase and the straight-line spin chirality can be applied to other systems is an open question.

At present, the $J_3$-dominant situation is realized only in the uniform kagome antiferromagnet BaCu$_3$V$_2$O$_8$(OD)$_2$ \cite{CoplanarOct_Boldrin_prl_18} but not in so-far reported breathing-kagome magnets \cite{GdRuAl_Hirschberger_natcom_19, DQVOF_Orain_prl_17, DQVOF_Clark_prl_13, LiAMoO_Haraguchi_prb_15, LiAMoO_Sharbaf_prl_18, PbOFCu_Zhang_ChemComm_20, YbNiGe_Takahashi_jpsj_20, FeSn_Tanaka_prb_20, CaCrO_Balz_natphys_16, Dy3Ru4Al12_Gao_prb_19}. When a relevant magnet is synthesized in the future, the asymmetric spin-wave dispersion could be observed in the neutron scattering experiment as a signature of the chiral phase. In transport measurements, although controlling the wave length of the spin wave to detect the ${\bf q}$-dependent nonreciprocity might be challenging, other aspects of the scalar spin chirality and the associated geometric phase may possibly be detectable, similarly to the magnon Hall effect of SOC origin \cite{MHE_Katsura_prl_10, MHE_Onose_science_10, MHE_Matsumoto_prl_11}. Also, since the thermal current is related to the scalar spin chirality ${\bf S}_i\cdot({\bf S}_j \times {\bf S}_k)$, the chiral transition might affect the thermal transport at finite temperatures, as in the case of the spin current relevant to the vector spin chirality ${\bf S}_i \times {\bf S}_j$ \cite{SpinDyn_Kawasaki_67, trans-sq_AK_prb_19, Z2_AK_prl_20, trans-cubic_A_prb_22}. Although these issues should be clarified in our future work, we believe that this work will promote the understanding of fundamental aspects of the spin dynamics and the associated transport phenomena in noncoplanar spin textures. 
 
\begin{acknowledgments}
The authors thank S. Hayami, M. Gohlke, and T. Shimokawa for useful comments and discussions. We are thankful to ISSP, the University of Tokyo and YITP, Kyoto University for providing us with CPU time. This work is supported by JSPS KAKENHI Grant No. JP21K03469, JP23H00257, and JP24K00572.
\end{acknowledgments}

\appendix
\section{Details of the spin-wave expansion}
For the spin configuration in the $\chi^{\rm T}>0$ state given by Eq. (\ref{eq:spin_conf}), we derive Eq. (\ref{eq:Bloch_linear}). Since the quantization axis at each sublattice $\tilde{S}^z_{\mu l}$ is chosen in the $\overline{\bf S}_{\mu l}$ direction and thus, depends on the sublattices, we first introduce the spin-space rotation to relate these spin vectors. In general, when a vector ${\bf r}$ is rotated by angle $\theta$ in the anti-clockwise direction around a rotation axis $\hat{n}$, the vector after rotation ${\bf r}^\prime$ is given by the following Rodrigues' rotation formula
\begin{equation}\label{eq:Rodrigues}
{\bf r}^\prime = {\bf r} \cos\theta +(1-\cos\theta) ({\bf r}\cdot \hat{n})\hat{n} + (\hat{n}\times{\bf r}) \sin\theta .
\end{equation} 
For the 12 sublattices characterized by the set of $\mu =$ I$-$IV and $l=$1$-$3 [see Fig. \ref{fig1}(a)], we introduce right-handed orthonormal vectors $(\hat{x}_{\mu l}, \hat{y}_{\mu l}, \hat{z}_{\mu l})$ defined in the laboratory frame. By using Eq. (\ref{eq:Rodrigues}), we rotate $\hat{z}_{\mu l}$ such that the rotated vector $\hat{z}^\prime_{\mu l}$ is aligned to $\overline{\bf S}_{\mu l}$. After the rotation, the remaining two vectors $\hat{x}^\prime_{\mu l}$ and $\hat{y}^\prime_{\mu l}$, respectively, determine the directions of the $x$ and $y$ components of the spin fluctuation $\delta \tilde{S}^x_{\mu l}$ and $\delta \tilde{S}^y_{\mu l}$ at each site. In this procedure, the rotation axis $\hat{n}$ and angle $\theta$ are determined in the following way. When we rotate $\hat{P}_1$ by the angle $\theta$ around $\hat{n}=(\hat{P}_2-\hat{P}_3)/\sqrt{2}$, we have
\begin{eqnarray}\label{eq:P1}
\hat{P}_1^\prime &=&  \hat{P}_1 \cos\theta +(1-\cos\theta) (\hat{P}_1 \cdot \hat{n})\hat{n} + (\hat{n}\times \hat{P}_1) \sin\theta \nonumber\\
&=& \hat{P}_1 \cos\theta - \frac{1}{\sqrt{2}} (\hat{P}_2 + \hat{P}_3) \sin\theta,
\end{eqnarray}
where the relations $\hat{P}_1 \cdot \hat{n}= 0$ and $\hat{n}\times \hat{P}_1=\frac{1}{\sqrt{2}}(\hat{P}_2\times\hat{P}_1 - \hat{P}_3 \times \hat{P}_1)= \frac{1}{\sqrt{2}}(-\hat{P}_3-\hat{P}_2)$ have been used. By comparing Eq. (\ref{eq:P1}) with $\overline{\bf S}_{{\rm I} 1} = c \hat{P}_1 + s (\hat{P}_2+\hat{P}_3)$ and $\overline{\bf S}_{{\rm IV} 1} = c \hat{P}_1 - s (\hat{P}_2+\hat{P}_3)$ in Eq. (\ref{eq:spin_conf}), one finds that $\hat{P}_1^\prime =  \overline{\bf S}_{{\rm I} 1}$ for $\theta=-\phi$ and $\hat{P}_1^\prime = \overline{\bf S}_{{\rm IV} 1}$ for $\theta=\phi$, where $\phi$ satisfies $\cos\phi = c = \frac{1}{\sqrt{1+2\varepsilon^2}}$ and $\sin\phi = s \sqrt{2} = \frac{\varepsilon \sqrt{2}}{\sqrt{1+2\varepsilon^2}}$. In the same manner, by rotating $-\hat{P}_1$ by the angles $\theta=\phi$ and $-\phi$ around $\hat{n}=(\hat{P}_2+\hat{P}_3)/\sqrt{2}$, we obtain $\overline{\bf S}_{{\rm II} 1} = -c\hat{P}_1 - s (\hat{P}_2-\hat{P}_3)$ and $ \overline{\bf S}_{{\rm III} 1} = -c\hat{P}_1 + s (\hat{P}_2-\hat{P}_3)$, respectively. The orthonormal vectors $(\hat{x}_{\mu l}, \hat{y}_{\mu l}, \hat{z}_{\mu l})$ and associated $\hat{n}$ and $\theta$ for the 12 sublattices are summarized in Table \ref{table1}. 

\begin{table}[b]\label{table1}
\caption{ The orthonormal vectors $(\hat{x}_{\mu l}, \hat{y}_{\mu l}, \hat{z}_{\mu l})$ and the rotation parameters $\hat{n}$ and $\theta$ for the sublattice characterized by $\mu$ and $l$. \label{table1}}
\begin{tabular}{|c|c|c|c|}
\hline
$\mu$  $l$ & $(\hat{x}_{\mu l}, \hat{y}_{\mu l}, \hat{z}_{\mu l})$ & $\hat{n}$ & $\theta$ \\
\hline
I 1 &  $(\hat{P}_2,\hat{P}_3,\hat{P}_1)$ & $\frac{1}{\sqrt{2}}(\hat{P}_2-\hat{P}_3)$ & $-\phi$ \\
\hline
II 1 & $(\hat{P}_2,-\hat{P}_3,-\hat{P}_1)$ & $\frac{1}{\sqrt{2}}(\hat{P}_2+\hat{P}_3)$ & $\phi$ \\
\hline
III 1 & $(\hat{P}_2,-\hat{P}_3,-\hat{P}_1)$ & $\frac{1}{\sqrt{2}}(\hat{P}_2+\hat{P}_3)$ & $-\phi$\\
\hline
IV 1 & $(\hat{P}_2,\hat{P}_3, \hat{P}_1)$ & $\frac{1}{\sqrt{2}}(\hat{P}_2-\hat{P}_3)$ & $\phi$ \\
\hline\hline
I 2 &  $(\hat{P}_3,-\hat{P}_1,-\hat{P}_2)$ & $\frac{1}{\sqrt{2}}(\hat{P}_3+\hat{P}_1)$ & $-\phi$ \\
\hline
II 2 & $(\hat{P}_3,\hat{P}_1,\hat{P}_2)$ & $\frac{1}{\sqrt{2}}(\hat{P}_3-\hat{P}_1)$ & $-\phi$ \\
\hline
III 2 & $(\hat{P}_3,-\hat{P}_1,-\hat{P}_2)$ & $\frac{1}{\sqrt{2}}(\hat{P}_3+\hat{P}_1)$ & $\phi$\\
\hline
IV 2 & $(\hat{P}_3,\hat{P}_1,\hat{P}_2)$ & $\frac{1}{\sqrt{2}}(\hat{P}_3-\hat{P}_1)$ & $\phi$ \\
\hline\hline
I 3 &  $(\hat{P}_1,-\hat{P}_2,-\hat{P}_3)$ & $\frac{1}{\sqrt{2}}(\hat{P}_1+\hat{P}_2)$ & $\phi$ \\
\hline
II 3 & $(\hat{P}_1,-\hat{P}_2,-\hat{P}_3)$ & $\frac{1}{\sqrt{2}}(\hat{P}_1+\hat{P}_2)$ & $-\phi$ \\
\hline
III 3 & $(\hat{P}_1,\hat{P}_2,\hat{P}_3)$ & $\frac{1}{\sqrt{2}}(\hat{P}_1-\hat{P}_2)$ & $-\phi$\\
\hline
IV 3 & $(\hat{P}_1,\hat{P}_2,\hat{P}_3)$ & $\frac{1}{\sqrt{2}}(\hat{P}_1-\hat{P}_2)$ & $\phi$ \\
\hline
\end{tabular}
\end{table} 

By using the triad after the rotation $(\hat{x}^\prime_{\mu l}, \hat{y}^\prime_{\mu l}, \hat{z}^\prime_{\mu l})$, one can express ${\bf S}_{\mu l} = \overline{\bf S}_{\mu l} + \delta {\bf S}_{\mu l}$ as ${\bf S}_{\mu l} =\hat{z}^\prime_{\mu l}+\delta \tilde{S}^x_{\mu l} \, \hat{x}^\prime_{\mu l}+ \delta \tilde{S}^y_{\mu l} \, \hat{y}^\prime_{\mu l}$. Note that $\hat{z}^\prime_{\mu l}$ is aligned to $\overline{\bf S}_{\mu l}$ with $|\overline{\bf S}_{\mu l}|=1$. Since $\hat{x}^\prime_{\mu l}$ and $\hat{y}^\prime_{\mu l}$ consist of the laboratory-frame orthonormal vectors $\hat{P}_1$, $\hat{P}_2$, and $\hat{P}_3$, $\delta {\bf S}_{\mu l}$ is expressed as 
 \begin{equation}\label{eq:fluc_compact}
 \delta {\bf S}_{\mu l} =  \delta \tilde{S}^x_{\mu l} \sum_{k =1}^3 c^{x,k}_{\mu l} \hat{P}_{k} + \delta \tilde{S}^y_{\mu l} \sum_{k =1}^3 c^{y,k}_{\mu l} \hat{P}_{k}, 
 \end{equation}
where the coefficients $c^{x,k}_{\mu l}$ and $c^{y,k}_{\mu l}$ are given in Table \ref{table2}. For later convenience, we introduce the Fourier transform $\delta \tilde{S}^{x (y)}_{\mu l, {\bf q}}  = \sum_{i \in (\mu, l)} \delta \tilde{S}^{x(y)}_{\mu l} e^{-i {\bf q}\cdot{\bf r}_i}$, where $\sum_{i \in (\mu, l)}$ denotes the summation over the sites belonging to the sublattice $(\mu, l)$. Then, Eq. (\ref{eq:fluc_compact}) can be rewritten as
 \begin{equation}\label{eq:fluc_compact_FT}
 \delta {\bf S}_{{\mu l},{\bf q}} =  \delta \tilde{S}^x_{{\mu l},{\bf q}} \sum_{k =1}^3 c^{x,k}_{\mu l} \hat{P}_{k} + \delta \tilde{S}^y_{{\mu l},{\bf q}} \sum_{k =1}^3 c^{y,k}_{\mu l} \hat{P}_{k}. 
 \end{equation}

\begin{table}[b]\label{table2}
\caption{ The coefficients $c^{x,k}_{\mu l}$ and $c^{y,k}_{\mu l}$ in Eq. (\ref{eq:fluc_compact}). \label{table2}}
\begin{tabular}{|c|c|c|c|}
\hline
$\mu$  $l$ & $k$ & $c^{x,k}_{\mu l}$ & $c^{y,k}_{\mu l}$  \\
\hline
  & 1 & $-s$ & $-s$ \\
I 1 &  2 &$(1+c)/2$ & $(-1+c)/2$  \\
 &  3 &$(-1+c)/2$ & $(1+c)/2$ \\
\hline
  & 1 & $-s$ & $-s$ \\
II 1 &  2 &$(1+c)/2$ & $(-1+c)/2$  \\
 &  3 &$-(-1+c)/2$ & $-(1+c)/2$ \\
\hline
  & 1 & $s$ & $s$ \\
III 1 &  2 &$(1+c)/2$ & $(-1+c)/2$  \\
 &  3 &$-(-1+c)/2$ & $-(1+c)/2$ \\
\hline
  & 1 & $s$ & $s$ \\
IV 1 &  2 &$(1+c)/2$ & $(-1+c)/2$  \\
 &  3 &$(-1+c)/2$ & $(1+c)/2$ \\
\hline\hline
  & 1 & $-(-1+c)/2$ & $-(1+c)/2$ \\
I 2 &  2 &$s$ & $s$  \\
 &  3 &$(1+c)/2$ & $(-1+c)/2$ \\
\hline
  & 1 & $(-1+c)/2$ & $(1+c)/2$ \\
II 2 &  2 &$-s$ & $-s$  \\
 &  3 &$(1+c)/2$ & $(-1+c)/2$ \\
\hline
  & 1 & $-(-1+c)/2$ & $-(1+c)/2$ \\
III 2 &  2 &$-s$ & $-s$  \\
 &  3 &$(1+c)/2$ & $(-1+c)/2$ \\
\hline
  & 1 & $(-1+c)/2$ & $(1+c)/2$ \\
IV 2 &  2 &$s$ & $s$  \\
 &  3 &$(1+c)/2$ & $(-1+c)/2$ \\
\hline\hline
  & 1 & $(1+c)/2$ & $(-1+c)/2$ \\
I 3 &  2 &$-(-1+c)/2$ & $-(1+c)/2$  \\
 &  3 &$-s$ & $-s$ \\
\hline
  & 1 & $(1+c)/2$ & $(-1+c)/2$ \\
II 3 &  2 &$-(-1+c)/2$ & $-(1+c)/2$  \\
 &  3 &$s$ & $s$ \\
\hline
  & 1 & $(1+c)/2$ & $(-1+c)/2$ \\
III 3 &  2 &$(-1+c)/2$ & $(1+c)/2$  \\
 &  3 &$-s$ & $-s$ \\
\hline
  & 1 & $(1+c)/2$ & $(-1+c)/2$ \\
IV 3 &  2 &$(-1+c)/2$ & $(1+c)/2$  \\
 &  3 &$s$ & $s$ \\
\hline
\end{tabular}
\end{table} 

Now, we derive Eq. (\ref{eq:Bloch_linear}). By substituting ${\bf S}_{\mu l} = \overline{\bf S}_{\mu l} + \delta {\bf S}_{\mu l}$ into Eq. (\ref{eq:Bloch}), we obtain the linearized equation of motion as 
\begin{equation}
\frac{d}{dt}\delta {\bf S}_{\mu l} = -\overline{\bf H}_{\mu l} \times \delta{\bf S}_{\mu l} - \delta {\bf H}_{\mu l}\times \overline{\bf S}_{\mu l}, 
\end{equation}
where $\overline{\bf H}_{\mu l}$ and $\delta {\bf H}_{\mu l}$ represent the zeroth- and first-order components of ${\bf H}^{\rm eff}_{\mu l}$ with respect to $\delta {\bf S}_{\mu l}$. In the momentum space representation, it reads
\begin{equation}\label{eq:spin-dynamics}
\frac{d}{dt}\delta {\bf S}_{{\mu l}, {\bf q}} = -\overline{\bf H}_{\mu l} \times \delta{\bf S}_{{\mu l},{\bf q}} - \delta {\bf H}_{{\mu l},{\bf q}} \times \overline{\bf S}_{\mu l}. 
\end{equation}
Here, the zeroth-order local field $\overline{\bf H}_{\mu l}$ can straightforwardly be calculated as
\begin{eqnarray}\label{eq:Hlocal}
&& \mspace{-32mu}\left\{ \begin{array}{ll}
\overline{\bf H}_{{\rm I} 1} =  h \hat{P}_1 - h^\prime (\hat{P}_2 + \hat{P}_3) ,& \overline{\bf H}_{{\rm II} 1} = -h \hat{P}_1 + h^\prime (\hat{P}_2 -  \hat{P}_3) , \\
\overline{\bf H}_{{\rm III} 1} = -h \hat{P}_1 - h^\prime (\hat{P}_2 - \hat{P}_3) , & \overline{\bf H}_{{\rm IV} 1} = h \hat{P}_1 + h^\prime (\hat{P}_2 +  \hat{P}_3) , 
\end{array} \right . \nonumber\\
&&\mspace{-32mu} \left\{ \begin{array}{ll}
\overline{\bf H}_{{\rm I} 2} =  - h \hat{P}_2 + h^\prime (\hat{P}_1 - \hat{P}_3) , & \overline{\bf H}_{{\rm II} 2} =  h \hat{P}_2 -h^\prime (\hat{P}_1 + \hat{P}_3) , \\
\overline{\bf H}_{{\rm III} 2} =  - h \hat{P}_2 -h^\prime (\hat{P}_1 - \hat{P}_3) , &  \overline{\bf H}_{{\rm IV} 2} = h \hat{P}_2 + h^\prime (\hat{P}_1  + \hat{P}_3) , 
\end{array} \right . \nonumber\\
&& \mspace{-32mu} \left\{ \begin{array}{ll}
\overline{\bf H}_{{\rm I} 3} = - h \hat{P}_3+ h^\prime (\hat{P}_1 - \hat{P}_2) , & \overline{\bf H}_{{\rm II} 3} =  - h \hat{P}_3 -h^\prime (\hat{P}_1 -  \hat{P}_2)  , \\
\overline{\bf H}_{{\rm III} 3} = h \hat{P}_3 -h^\prime (\hat{P}_1 + \hat{P}_2),  &  \overline{\bf H}_{{\rm IV} 3} =  h \hat{P}_3+h^\prime ( \hat{P}_1 + \hat{P}_2 )  ,
\end{array} \right . \nonumber\\
&& h = 2 s J_1-2sJ_1^\prime +4 c J_3, \nonumber\\
&& h^\prime  =  (s-c) J_1 +  (s+c) J_1^\prime . 
\end{eqnarray}
As for the first-order contribution $\delta {\bf H}_{{\mu l},{\bf q}}$, by using $G_{i {\bf q}} = J_1 e^{i{\bf q}\cdot {\bf e}_i}$, $G^\prime_{i {\bf q}} = J_1^\prime e^{i{\bf q}\cdot {\bf e}_i}$, and $T_{i{\bf q}} = 2J_3\cos(2{\bf q}\cdot{\bf e}_i) $ in Eq. (\ref{eq:coefficient}), we have
\begin{eqnarray}\label{eq:H_fluc}
&&\mspace{-40mu} \delta {\bf H}_{{\mu 1},{\bf q}} =  \sum_{k=1}^3 \big[ G_{1{\bf q}} w^{1,k}_{\mu 1} + G^\ast_{3{\bf q}} w^{2,k}_{\mu 1} + G^{\prime \ast}_{1{\bf q}} w^{3,k}_{\mu 1} \nonumber\\
&&\mspace{-32mu} \qquad\qquad\, + G^\prime_{3{\bf q}}w^{4,k}_{\mu 1}+ T_{1{\bf q}} w^{5,k}_{\mu 1}+ T_{3{\bf q}} w^{6,k}_{\mu 1} \big] \hat{P}_k ,  \nonumber\\
&&\mspace{-40mu} \delta {\bf H}_{{\mu 2},{\bf q}} =  \sum_{k=1}^3 \big[ G^\ast_{1{\bf q}} w^{1,k}_{\mu 2} + G_{2{\bf q}} w^{2,k}_{\mu 2} + G^{\prime}_{1{\bf q}} w^{3,k}_{\mu 2} \nonumber\\
&&\mspace{-32mu} \qquad\qquad\, + G^{\prime \ast}_{2{\bf q}}w^{4,k}_{\mu 2}+ T_{1{\bf q}} w^{5,k}_{\mu 2}+ T_{2{\bf q}} w^{6,k}_{\mu 2} \big] \hat{P}_k ,  \nonumber\\
&&\mspace{-40mu} \delta {\bf H}_{{\mu 3},{\bf q}} =  \sum_{k=1}^3 \big[ G_{3{\bf q}} w^{1,k}_{\mu 3} + G^\ast_{2{\bf q}} w^{2,k}_{\mu 3} + G^{\prime \ast}_{3{\bf q}} w^{3,k}_{\mu 3} \nonumber\\
&&\mspace{-32mu} \qquad\qquad\, + G^\prime_{2{\bf q}}w^{4,k}_{\mu 3}+ T_{3{\bf q}} w^{5,k}_{\mu 3}+ T_{2{\bf q}} w^{6,k}_{\mu 3} \big] \hat{P}_k   
\end{eqnarray}
with the spin-fluctuation weight $w^{i,k}_{\mu l}$ shown in Table \ref{table3}. 

By substituting Eqs. (\ref{eq:fluc_compact_FT}), (\ref{eq:Hlocal}), and (\ref{eq:H_fluc}) into Eq. (\ref{eq:spin-dynamics}) and taking the basis change 
\begin{equation}\label{eq:basis}
\left(\begin{array}{c}
\delta \tilde{S}^x_{{\rm I}1,{\bf q}} \\
\delta \tilde{S}^x_{{\rm I}2,{\bf q}}\\
\delta \tilde{S}^x_{{\rm I}3,{\bf q}}\\
\vdots \\
\delta \tilde{S}^x_{{\rm IV}1,{\bf q}}\\
\delta \tilde{S}^x_{{\rm IV}2,{\bf q}}\\
\delta \tilde{S}^x_{{\rm IV}3,{\bf q}}\\
\\
\delta \tilde{S}^y_{{\rm I}1,{\bf q}} \\
\delta \tilde{S}^y_{{\rm I}2,{\bf q}}\\
\delta \tilde{S}^y_{{\rm I}3,{\bf q}}\\
\vdots \\
\delta \tilde{S}^y_{{\rm IV}1,{\bf q}}\\
\delta \tilde{S}^y_{{\rm IV}2,{\bf q}}\\
\delta \tilde{S}^y_{{\rm IV}3,{\bf q}}
\end{array}
\right) \rightarrow \mbox{\boldmath $\Phi$}_{\bf q}=\left(\begin{array}{c}
\delta \tilde{S}^+_{{\rm I}1,{\bf q}} \\
\delta \tilde{S}^+_{{\rm I}2,{\bf q}}\\
\delta \tilde{S}^+_{{\rm I}3,{\bf q}}\\
\vdots \\
\delta \tilde{S}^+_{{\rm IV}1,{\bf q}}\\
\delta \tilde{S}^+_{{\rm IV}2,{\bf q}}\\
\delta \tilde{S}^+_{{\rm IV}3,{\bf q}}\\
\\
\delta \tilde{S}^-_{{\rm I}1,{\bf q}} \\
\delta \tilde{S}^-_{{\rm I}2,{\bf q}}\\
\delta \tilde{S}^-_{{\rm I}3,{\bf q}}\\
\vdots \\
\delta \tilde{S}^-_{{\rm IV}1,{\bf q}}\\
\delta \tilde{S}^-_{{\rm IV}2,{\bf q}}\\
\delta \tilde{S}^-_{{\rm IV}3,{\bf q}}
\end{array}
\right) 
\end{equation}
with $\delta \tilde{S}^{\pm}_{{\mu l},{\bf q}}\equiv  \delta \tilde{S}^x_{{\mu l},{\bf q}} \pm i \delta \tilde {S}^y_{{\mu l},{\bf q}}$, we obtain Eq. (\ref{eq:Bloch_linear}). 

\section{Dynamical spin structure factor}
The information of the magnetic excitations including the spin waves can be captured by the dynamical spin structure factor $S({\bf q},\omega)= |\frac{1}{t_0}\int dt \, \frac{1}{N} \sum_i {\bf S}_i e^{-i({\bf q}\cdot {\bf r}_i-\omega t)}|^2$. Here, we explain how to calculate $S({\bf q},\omega)$ in the numerical simulation and in the spin-wave expansion. 
 
 In the numerical simulation, we prepare finite-temperature equilibrium spin configurations in the Monte Carlo (MC) simulation and then, numerically integrate Eq. (\ref{eq:Bloch}) with the MC snapshots as the initial state. In our MC simulation for the total number of spins $N=3L^2$ with $L=288$, we perform $10^5$ MC sweeps for thermalization and then, pick up MC snapshots, where our one MC sweep consists of one heatbath sweep and successive 10 over-relaxation sweeps. By using the second order symplectic method \cite{Symplectic_Krech_98}, we integrate Eq. (\ref{eq:Bloch}) up to $t_0=1000/J_1$ with the time step $\delta t = 0.01/J_1$ to obtain $S({\bf q}, \omega)$. To have sufficiently high resolution in the ${\bf q}$ and $\omega$ plane, we use the large values of $L=288$ and $t_0=1000/J_1$. In general, when we increase $t_0$ with $L$ being fixed, relevant spin-wave modes become clearer in $S({\bf q}, \omega)$, which has actually been confirmed in this work. Since it is difficult to check the convergence at each point in the whole ${\bf q}$ and $\omega$ space as associated modes are generally incommensurate with both $L$ and $t_0$, we have simply checked that the spectrum obtained for $t_0=1000/J_1$ is essentially the same as those for smaller values of $t_0$. As for the $L$ dependence, in the present two-dimensional Heisenberg spin system, a long-range magnetic order is not allowed at any finite temperatures, so that with increasing $L$, the finite-temperature spin structure factor should gradually decrease. Nevertheless, we have also checked that the finite-temperature $S({\bf q}, \omega)$'s obtained for $L=144$ and 288 are qualitatively unchanged and that they are consistent with the analytical result obtained in the spin-wave expansion which corresponds to the $L, t_0 \rightarrow \infty$ limit at $T=0$, as is demonstrated below.

In the spin-wave expansion, the spin dynamics is governed by Eq. (\ref{eq:Bloch_linear}) in which the eigen value of $H_{\bf q}$, i.e., $\omega=E({\bf q})$, gives the spin-wave dispersion. Since as shown in Eq. (\ref{eq:basis}), the associated eigen vector $\mbox{\boldmath $\Phi$}_{\bf q}(\omega)$ describes the spin fluctuations $\delta \tilde{S}_{{\mu l},{\bf q}}^\pm$ relevant to the eigen mode, the dynamical spin structure factor $S({\bf q},\omega)$ can be calculated from $\mbox{\boldmath $\Phi$}_{\bf q}(\omega)$. Due to the orthogonality $\overline{\bf S}_i \perp \delta {\bf S}_i$, we have
\begin{eqnarray}
&& S({\bf q},\omega) =  \Big|\frac{1}{t_0} \int dt \, \frac{1}{N}\sum_i (\overline{\bf S}_i + \delta {\bf S}_i)  e^{-i({\bf q}\cdot {\bf r}_i-\omega t)} \Big|^2\nonumber\\
&&\quad =\delta_{\omega 0}\Big|\frac{1}{N}\sum_i  \overline{\bf S}_i  e^{-i{\bf q}\cdot {\bf r}_i}\Big|^2 + \Big|\frac{1}{t_0}\int dt \, \frac{1}{N}\sum_i \delta {\bf S}_i  e^{-i({\bf q}\cdot {\bf r}_i-\omega t)}\Big|^2  \nonumber\\
&&\quad = S({\bf q}, 0) + \Big| \frac{1}{t_0} \int dt \, \delta {\bf S}_{\bf q} e^{i\omega t}\Big|^2.  \nonumber
\end{eqnarray}
Since the first term corresponds to the static spin structure factor, the problem is reduced to calculate the second term $\big| \frac{1}{t_0}\int dt \, \delta {\bf S}_{\bf q} e^{i\omega t}\big|^2$ which in the present system, can be written as
\begin{eqnarray}\label{eq:Sq_omega}
&& \Big| \sum_{\mu, l} \frac{1}{t_0} \int dt \, \delta {\bf S}_{\mu l, {\bf q}} e^{i\omega t}\Big|^2  \nonumber\\
&& =  \Big|\sum_{\mu, l} \frac{1}{t_0}\int dt \big( \delta \tilde{S}^x_{{\mu l},{\bf q}} \sum_{k =1}^3 c^{x,k}_{\mu l} \hat{P}_{k} + \delta \tilde{S}^y_{{\mu l},{\bf q}} \sum_{k =1}^3 c^{y,k}_{\mu l} \hat{P}_{k} \big) e^{i\omega t} \Big|^2 \nonumber\\
&& =  \sum_{k=1}^3 \Big|\sum_{\mu, l} \frac{1}{t_0} \int dt \big( c^{x,k}_{\mu l} \delta \tilde{S}^x_{{\mu l},{\bf q}} + c^{y,k}_{\mu l}\delta \tilde{S}^y_{{\mu l},{\bf q}} \big) e^{i\omega t} \Big|^2 \nonumber\\
&& =  \sum_{k=1}^3 \Big|\sum_{\mu, l}\Big( \frac{c^{x,k}_{\mu l} -i c^{y,k}_{\mu l}}{2}\delta \tilde{S}^+_{{\mu l},{\bf q}}(\omega) +\frac{c^{x,k}_{\mu l} +i c^{y,k}_{\mu l}}{2} \delta \tilde{S}^-_{{\mu l},{\bf q}} (\omega) \Big) \Big|^2 \nonumber\\
\end{eqnarray}
where Eq. (\ref{eq:fluc_compact_FT}) has been used. From the eigen vector $\mbox{\boldmath $\Phi$}_{\bf q}(\omega)$ having $\delta \tilde{S}^\pm_{{\mu l},{\bf q}}(\omega)$ in its components, one can calculate Eq. (\ref{eq:Sq_omega}) for each eigen mode $\omega = E({\bf q})$. 

\begin{figure}[t]
\includegraphics[scale=0.5]{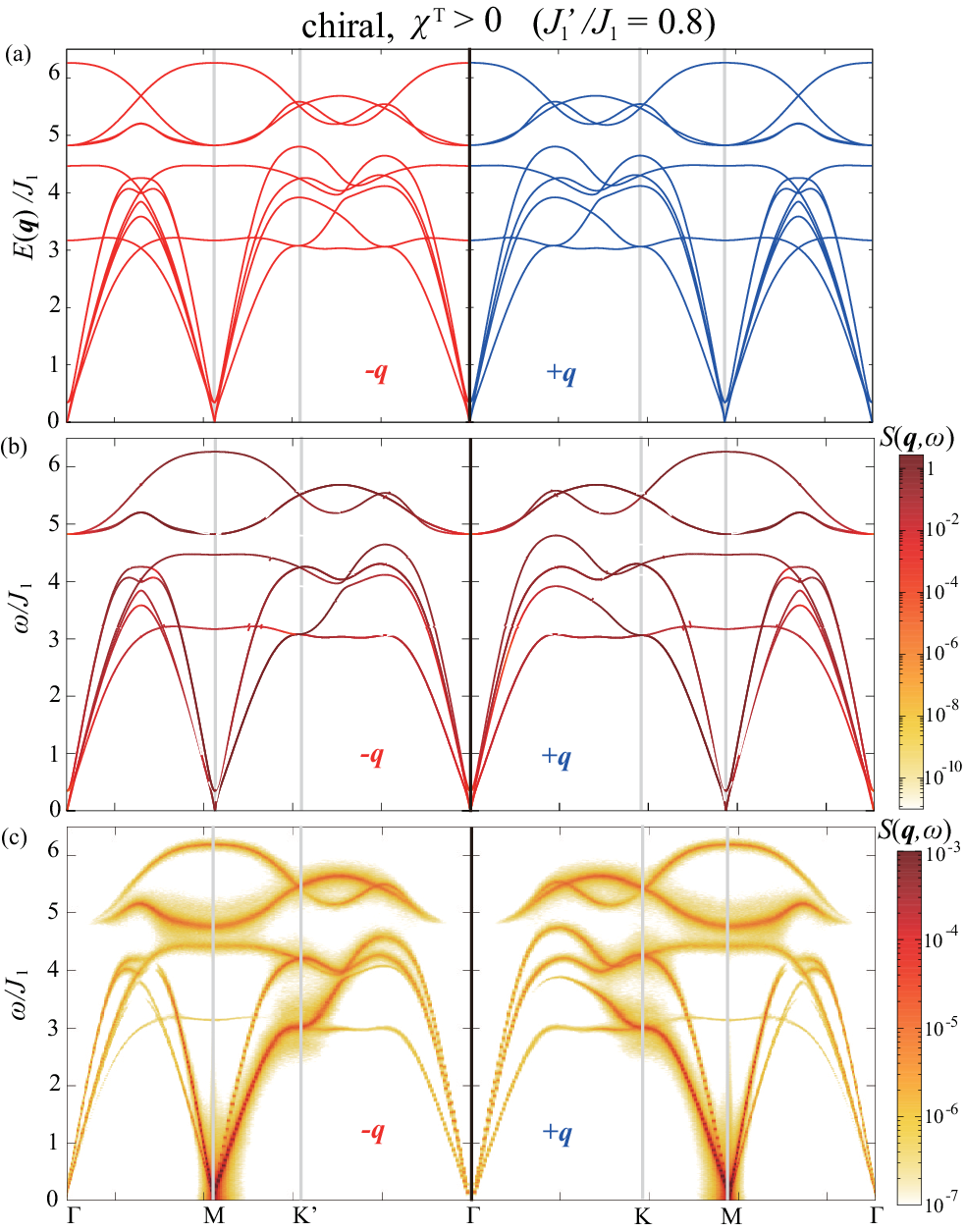}
\caption{ The excitation spectrum in the $\chi^{\rm T}>0$ phase for $J_1^\prime/J_1=0.8$. (a) The spin-wave dispersion and (b) dynamical spin structure factor $S({\bf q}, \omega)$ obtained in the spin-wave expansion. (c) $S({\bf q},\omega)$ obtained in the numerical simulation at $T/J_1=0.04$. In each figure, the left and right halves correspond to the red and blue trajectories in Fig. \ref{fig1} (b). (c) is obtained in the same numerical analysis as that for Fig. \ref{fig3}. \label{fig:SM1}}
\end{figure}

Figure \ref{fig:SM1} shows (a) the spin-wave dispersion $E({\bf q})$ and (b) and (c) the dynamical spin structure factors $S({\bf q},\omega)$ obtained in the $\chi^{\rm T}>0$ phase for $J_1^\prime/J_1=0.8$. Here, (a) and (b) are the results of the spin-wave expansion, whereas (c) is the result of the numerical simulation at $T/J_1=0.04$. Figure \ref{fig:SM1} (b) is obtained by plotting the spin-wave dispersion shown in (a) with the intensity calculated from Eq. (\ref{eq:Sq_omega}). One can see from Figs. \ref{fig:SM1} (b) and (c) that the analytical and numerical results are consistent with each other, supporting the validity of the spin-wave expansion. 

By comparing (a) with (b), one notices that some spin-wave branches disappear in $S({\bf q}, \omega)$. Such a difference between $E({\bf q})$ and $S({\bf q}, \omega)$ is generally possible, as they are different physical quantities. In particular, in the present system, the quantization axes differ site to site, and as a result, the eigen mode is a complicated superposition of the 12-sublattice spin fluctuations in three-dimensional spin space, as indicated by Eq. (\ref{eq:fluc_compact_FT}). Although $S({\bf q}, \omega)$ is not exactly the same as $E({\bf q})$, the asymmetry in $E({\bf q})$ can be seen even in the dynamical spin structure factor $S({\bf q}, \omega)$.

\clearpage
\begin{table}[htbp]\label{table3}
\caption{ The spin-fluctuation weight $w^{i,k}_{\mu l}$ in Eq. (\ref{eq:H_fluc}). \label{table3}}
\resizebox{\textwidth}{!}{
\begin{tabular}{|c|c|c|c|c|c|c|c|}
\hline
$\mu$  $l$ & $k$ & $w^{1,k}_{\mu l}$ & $w^{2,k}_{\mu l}$ & $w^{3,k}_{\mu l}$ & $w^{4,k}_{\mu l}$ & $w^{5,k}_{\mu l}$ & $w^{6,k}_{\mu l}$ \\
\hline
 & & & & & & & \\
  & 1 & $\frac{c-1}{2}\delta \tilde{S}^x_{{\rm I}2}+\frac{c+1}{2}\delta \tilde{S}^y_{{\rm I}2}$ & $-\frac{c+1}{2}\delta \tilde{S}^x_{{\rm I}3}-\frac{c-1}{2}\delta \tilde{S}^y_{{\rm I}3}$ & $-\frac{c-1}{2}\delta \tilde{S}^x_{{\rm II}2}-\frac{c+1}{2}\delta \tilde{S}^y_{{\rm II}2}$ & $-\frac{c+1}{2}\delta \tilde{S}^x_{{\rm III}3}-\frac{c-1}{2}\delta \tilde{S}^y_{{\rm III}3}$ & $s\delta \tilde{S}^x_{{\rm II}1}+s\delta \tilde{S}^y_{{\rm II}1}$ & $-s\delta \tilde{S}^x_{{\rm III}1}-s\delta \tilde{S}^y_{{\rm III}1}$ \\
I 1 & 2 & $-s\delta \tilde{S}^x_{{\rm I}2}-s\delta \tilde{S}^y_{{\rm I}2}$ & $\frac{c-1}{2}\delta \tilde{S}^x_{{\rm I}3}+\frac{c+1}{2}\delta \tilde{S}^y_{{\rm I}3}$ & $s\delta \tilde{S}^x_{{\rm II}2}+s\delta \tilde{S}^y_{{\rm II}2}$ & $-\frac{c-1}{2}\delta \tilde{S}^x_{{\rm III}3}-\frac{c+1}{2}\delta \tilde{S}^y_{{\rm III}3}$ & $-\frac{c+1}{2}\delta \tilde{S}^x_{{\rm II}1}-\frac{c-1}{2}\delta \tilde{S}^y_{{\rm II}1}$ & $-\frac{c+1}{2}\delta \tilde{S}^x_{{\rm III}1}-\frac{c-1}{2}\delta \tilde{S}^y_{{\rm III}1}$ \\
 & 3 & $-\frac{c+1}{2}\delta \tilde{S}^x_{{\rm I}2}-\frac{c-1}{2}\delta \tilde{S}^y_{{\rm I}2}$ & $ s \tilde{S}^x_{{\rm I}3}+s\delta \tilde{S}^y_{{\rm I}3}$ & $-\frac{c+1}{2}\delta \tilde{S}^x_{{\rm II}2}-\frac{c-1}{2}\delta \tilde{S}^y_{{\rm II}2}$ & $s\delta \tilde{S}^x_{{\rm III}3}+s\delta \tilde{S}^y_{{\rm III}3}$ & $\frac{c-1}{2}\delta \tilde{S}^x_{{\rm II}1}+\frac{c+1}{2}\delta \tilde{S}^y_{{\rm II}1}$ & $\frac{c-1}{2}\delta \tilde{S}^x_{{\rm III}1}+\frac{c+1}{2}\delta \tilde{S}^y_{{\rm III}1}$ \\
 & & & & & & & \\ 
\hline
 & & & & & & & \\
   & 1 & $-\frac{c-1}{2}\delta \tilde{S}^x_{{\rm II}2}-\frac{c+1}{2}\delta \tilde{S}^y_{{\rm II}2}$ & $-\frac{c+1}{2}\delta \tilde{S}^x_{{\rm II}3}-\frac{c-1}{2}\delta \tilde{S}^y_{{\rm II}3}$ & $\frac{c-1}{2}\delta \tilde{S}^x_{{\rm I}2}+\frac{c+1}{2}\delta \tilde{S}^y_{{\rm I}2}$ & $-\frac{c+1}{2}\delta \tilde{S}^x_{{\rm IV}3}-\frac{c-1}{2}\delta \tilde{S}^y_{{\rm IV}3}$ & $s\delta \tilde{S}^x_{{\rm I}1}+s\delta \tilde{S}^y_{{\rm I}1}$ & $-s\delta \tilde{S}^x_{{\rm IV}1}-s\delta \tilde{S}^y_{{\rm IV}1}$ \\ 
II 1 & 2 & $s\delta \tilde{S}^x_{{\rm II}2}+s\delta \tilde{S}^y_{{\rm II}2}$ & $\frac{c-1}{2}\delta \tilde{S}^x_{{\rm II}3}+\frac{c+1}{2}\delta \tilde{S}^y_{{\rm II}3}$ & $-s\delta \tilde{S}^x_{{\rm I}2}-s\delta \tilde{S}^y_{{\rm I}2}$ & $-\frac{c-1}{2}\delta \tilde{S}^x_{{\rm IV}3}-\frac{c+1}{2}\delta \tilde{S}^y_{{\rm IV}3}$ & $-\frac{c+1}{2}\delta \tilde{S}^x_{{\rm I}1}-\frac{c-1}{2}\delta \tilde{S}^y_{{\rm I}1}$ & $-\frac{c+1}{2}\delta \tilde{S}^x_{{\rm IV}1}-\frac{c-1}{2}\delta \tilde{S}^y_{{\rm IV}1}$ \\
 & 3 & $-\frac{c+1}{2}\delta \tilde{S}^x_{{\rm II}2}-\frac{c-1}{2}\delta \tilde{S}^y_{{\rm II}2}$ & $ -s \tilde{S}^x_{{\rm II}3}-s\delta \tilde{S}^y_{{\rm II}3}$ & $-\frac{c+1}{2}\delta \tilde{S}^x_{{\rm I}2}-\frac{c-1}{2}\delta \tilde{S}^y_{{\rm I}2}$ & $-s\delta \tilde{S}^x_{{\rm IV}3}-s\delta \tilde{S}^y_{{\rm IV}3}$ & $-\frac{c-1}{2}\delta \tilde{S}^x_{{\rm I}1}-\frac{c+1}{2}\delta \tilde{S}^y_{{\rm I}1}$ & $-\frac{c-1}{2}\delta \tilde{S}^x_{{\rm IV}1}-\frac{c+1}{2}\delta \tilde{S}^y_{{\rm IV}1}$ \\
  & & & & & & & \\
   \hline
 & & & & & & & \\   
   & 1 & $\frac{c-1}{2}\delta \tilde{S}^x_{{\rm III}2}+\frac{c+1}{2}\delta \tilde{S}^y_{{\rm III}2}$ & $-\frac{c+1}{2}\delta \tilde{S}^x_{{\rm III}3}-\frac{c-1}{2}\delta \tilde{S}^y_{{\rm III}3}$ & $-\frac{c-1}{2}\delta \tilde{S}^x_{{\rm IV}2}-\frac{c+1}{2}\delta \tilde{S}^y_{{\rm IV}2}$ & $-\frac{c+1}{2}\delta \tilde{S}^x_{{\rm I}3}-\frac{c-1}{2}\delta \tilde{S}^y_{{\rm I}3}$ & $-s\delta \tilde{S}^x_{{\rm IV}1}-s\delta \tilde{S}^y_{{\rm IV}1}$ & $s\delta \tilde{S}^x_{{\rm I}1}+s\delta \tilde{S}^y_{{\rm I}1}$ \\
III 1 & 2 & $s\delta \tilde{S}^x_{{\rm III}2}+s\delta \tilde{S}^y_{{\rm III}2}$ & $-\frac{c-1}{2}\delta \tilde{S}^x_{{\rm III}3}-\frac{c+1}{2}\delta \tilde{S}^y_{{\rm III}3}$ & $-s\delta \tilde{S}^x_{{\rm IV}2}-s\delta \tilde{S}^y_{{\rm IV}2}$ & $\frac{c-1}{2}\delta \tilde{S}^x_{{\rm I}3}+\frac{c+1}{2}\delta \tilde{S}^y_{{\rm I}3}$ & $-\frac{c+1}{2}\delta \tilde{S}^x_{{\rm IV}1}-\frac{c-1}{2}\delta \tilde{S}^y_{{\rm IV}1}$ & $-\frac{c+1}{2}\delta \tilde{S}^x_{{\rm I}1}-\frac{c-1}{2}\delta \tilde{S}^y_{{\rm I}1}$ \\
 & 3 & $-\frac{c+1}{2}\delta \tilde{S}^x_{{\rm III}2}-\frac{c-1}{2}\delta \tilde{S}^y_{{\rm III}2}$ & $ s \tilde{S}^x_{{\rm III}3}+s\delta \tilde{S}^y_{{\rm III}3}$ & $-\frac{c+1}{2}\delta \tilde{S}^x_{{\rm IV}2}-\frac{c-1}{2}\delta \tilde{S}^y_{{\rm IV}2}$ & $s\delta \tilde{S}^x_{{\rm I}3}+s\delta \tilde{S}^y_{{\rm I}3}$ & $-\frac{c-1}{2}\delta \tilde{S}^x_{{\rm IV}1}-\frac{c+1}{2}\delta \tilde{S}^y_{{\rm IV}1}$ & $-\frac{c-1}{2}\delta \tilde{S}^x_{{\rm I}1}-\frac{c+1}{2}\delta \tilde{S}^y_{{\rm I}1}$ \\
 & & & & & & & \\
 \hline
 & & & & & & & \\
   & 1 & $-\frac{c-1}{2}\delta \tilde{S}^x_{{\rm IV}2}-\frac{c+1}{2}\delta \tilde{S}^y_{{\rm IV}2}$ & $-\frac{c+1}{2}\delta \tilde{S}^x_{{\rm IV}3}-\frac{c-1}{2}\delta \tilde{S}^y_{{\rm IV}3}$ & $\frac{c-1}{2}\delta \tilde{S}^x_{{\rm III}2}+\frac{c+1}{2}\delta \tilde{S}^y_{{\rm III}2}$ & $-\frac{c+1}{2}\delta \tilde{S}^x_{{\rm II}3}-\frac{c-1}{2}\delta \tilde{S}^y_{{\rm II}3}$ & $-s\delta \tilde{S}^x_{{\rm III}1}-s\delta \tilde{S}^y_{{\rm III}1}$ & $s\delta \tilde{S}^x_{{\rm II}1}+s\delta \tilde{S}^y_{{\rm II}1}$ \\
IV 1 & 2 & $-s\delta \tilde{S}^x_{{\rm IV}2}-s\delta \tilde{S}^y_{{\rm IV}2}$ & $-\frac{c-1}{2}\delta \tilde{S}^x_{{\rm IV}3}-\frac{c+1}{2}\delta \tilde{S}^y_{{\rm IV}3}$ & $s\delta \tilde{S}^x_{{\rm III}2}+s\delta \tilde{S}^y_{{\rm III}2}$ & $\frac{c-1}{2}\delta \tilde{S}^x_{{\rm II}3}+\frac{c+1}{2}\delta \tilde{S}^y_{{\rm II}3}$ & $-\frac{c+1}{2}\delta \tilde{S}^x_{{\rm III}1}-\frac{c-1}{2}\delta \tilde{S}^y_{{\rm III}1}$ & $-\frac{c+1}{2}\delta \tilde{S}^x_{{\rm II}1}-\frac{c-1}{2}\delta \tilde{S}^y_{{\rm II}1}$ \\
 & 3 & $-\frac{c+1}{2}\delta \tilde{S}^x_{{\rm IV}2}-\frac{c-1}{2}\delta \tilde{S}^y_{{\rm IV}2}$ & $ -s \tilde{S}^x_{{\rm IV}3}-s\delta \tilde{S}^y_{{\rm IV}3}$ & $-\frac{c+1}{2}\delta \tilde{S}^x_{{\rm III}2}-\frac{c-1}{2}\delta \tilde{S}^y_{{\rm III}2}$ & $-s\delta \tilde{S}^x_{{\rm II}3}-s\delta \tilde{S}^y_{{\rm II}3}$ & $\frac{c-1}{2}\delta \tilde{S}^x_{{\rm III}1}+\frac{c+1}{2}\delta \tilde{S}^y_{{\rm III}1}$ & $\frac{c-1}{2}\delta \tilde{S}^x_{{\rm II}1}+\frac{c+1}{2}\delta \tilde{S}^y_{{\rm II}1}$ \\
 & & & & & & & \\
 \hline \hline
 & & & & & & & \\
   & 1 & $s\delta \tilde{S}^x_{{\rm I}1}+s\delta \tilde{S}^y_{{\rm I}1}$ & $-\frac{c+1}{2}\delta \tilde{S}^x_{{\rm I}3}-\frac{c-1}{2}\delta \tilde{S}^y_{{\rm I}3}$ & $s \delta \tilde{S}^x_{{\rm II}1}+s\delta \tilde{S}^y_{{\rm II}1}$ & $-\frac{c+1}{2}\delta \tilde{S}^x_{{\rm IV}3}-\frac{c-1}{2}\delta \tilde{S}^y_{{\rm IV}3}$ & $-\frac{c-1}{2}\delta \tilde{S}^x_{{\rm II}2}-\frac{c+1}{2}\delta \tilde{S}^y_{{\rm II}2}$ & $-\frac{c-1}{2}\delta \tilde{S}^x_{{\rm IV}2}-\frac{c+1}{2}\delta \tilde{S}^y_{{\rm IV}2}$ \\
I 2   & 2 & $-\frac{c+1}{2}\delta \tilde{S}^x_{{\rm I}1}-\frac{c-1}{2}\delta \tilde{S}^y_{{\rm I}1}$ & $\frac{c-1}{2}\delta \tilde{S}^x_{{\rm I}3}+\frac{c+1}{2}\delta \tilde{S}^y_{{\rm I}3}$ & $-\frac{c+1}{2}\delta \tilde{S}^x_{{\rm II}1}-\frac{c-1}{2}\delta \tilde{S}^y_{{\rm II}1}$ & $-\frac{c-1}{2}\delta \tilde{S}^x_{{\rm IV}3}-\frac{c+1}{2}\delta \tilde{S}^y_{{\rm IV}3}$ & $s\delta \tilde{S}^x_{{\rm II}2}+s\delta \tilde{S}^y_{{\rm II}2}$ & $-s\delta \tilde{S}^x_{{\rm IV}2}-s\delta \tilde{S}^y_{{\rm IV}2}$ \\
  & 3 & $-\frac{c-1}{2}\delta \tilde{S}^x_{{\rm I}1}-\frac{c+1}{2}\delta \tilde{S}^y_{{\rm I}1}$ & $s\delta \tilde{S}^x_{{\rm I}3}+s\delta \tilde{S}^y_{{\rm I}3}$ & $\frac{c-1}{2}\delta \tilde{S}^x_{{\rm II}1}+\frac{c+1}{2}\delta \tilde{S}^y_{{\rm II}1}$ & $-s\delta \tilde{S}^x_{{\rm IV}3}-s\delta \tilde{S}^y_{{\rm IV}3}$ & $-\frac{c+1}{2}\delta \tilde{S}^x_{{\rm II}2}-\frac{c-1}{2}\delta \tilde{S}^y_{{\rm II}2}$ & $-\frac{c+1}{2}\delta \tilde{S}^x_{{\rm IV}2}-\frac{c-1}{2}\delta \tilde{S}^y_{{\rm IV}2}$ \\
 & & & & & & & \\
 \hline
 & & & & & & & \\
   & 1 & $s\delta \tilde{S}^x_{{\rm II}1}+s\delta \tilde{S}^y_{{\rm II}1}$ & $-\frac{c+1}{2}\delta \tilde{S}^x_{{\rm II}3}-\frac{c-1}{2}\delta \tilde{S}^y_{{\rm II}3}$ & $s \delta \tilde{S}^x_{{\rm I}1}+s\delta \tilde{S}^y_{{\rm I}1}$ & $-\frac{c+1}{2}\delta \tilde{S}^x_{{\rm III}3}-\frac{c-1}{2}\delta \tilde{S}^y_{{\rm III}3}$ & $\frac{c-1}{2}\delta \tilde{S}^x_{{\rm I}2}+\frac{c+1}{2}\delta \tilde{S}^y_{{\rm I}2}$ & $\frac{c-1}{2}\delta \tilde{S}^x_{{\rm III}2}+\frac{c+1}{2}\delta \tilde{S}^y_{{\rm III}2}$ \\
II 2   & 2 & $-\frac{c+1}{2}\delta \tilde{S}^x_{{\rm II}1}-\frac{c-1}{2}\delta \tilde{S}^y_{{\rm II}1}$ & $\frac{c-1}{2}\delta \tilde{S}^x_{{\rm II}3}+\frac{c+1}{2}\delta \tilde{S}^y_{{\rm II}3}$ & $-\frac{c+1}{2}\delta \tilde{S}^x_{{\rm I}1}-\frac{c-1}{2}\delta \tilde{S}^y_{{\rm I}1}$ & $-\frac{c-1}{2}\delta \tilde{S}^x_{{\rm III}3}-\frac{c+1}{2}\delta \tilde{S}^y_{{\rm III}3}$ & $-s\delta \tilde{S}^x_{{\rm I}2}-s\delta \tilde{S}^y_{{\rm I}2}$ & $s\delta \tilde{S}^x_{{\rm III}2}+s\delta \tilde{S}^y_{{\rm III}2}$ \\
  & 3 & $\frac{c-1}{2}\delta \tilde{S}^x_{{\rm II}1}+\frac{c+1}{2}\delta \tilde{S}^y_{{\rm II}1}$ & $-s\delta \tilde{S}^x_{{\rm II}3}-s\delta \tilde{S}^y_{{\rm II}3}$ & $-\frac{c-1}{2}\delta \tilde{S}^x_{{\rm I}1}-\frac{c+1}{2}\delta \tilde{S}^y_{{\rm I}1}$ & $s\delta \tilde{S}^x_{{\rm III}3}+s\delta \tilde{S}^y_{{\rm III}3}$ & $-\frac{c+1}{2}\delta \tilde{S}^x_{{\rm I}2}-\frac{c-1}{2}\delta \tilde{S}^y_{{\rm I}2}$ & $-\frac{c+1}{2}\delta \tilde{S}^x_{{\rm III}2}-\frac{c-1}{2}\delta \tilde{S}^y_{{\rm III}2}$ \\
 & & & & & & & \\
 \hline
 & & & & & & & \\
   & 1 & $-s\delta \tilde{S}^x_{{\rm III}1}-s\delta \tilde{S}^y_{{\rm III}1}$ & $-\frac{c+1}{2}\delta \tilde{S}^x_{{\rm III}3}-\frac{c-1}{2}\delta \tilde{S}^y_{{\rm III}3}$ & $-s \delta \tilde{S}^x_{{\rm IV}1}-s\delta \tilde{S}^y_{{\rm IV}1}$ & $-\frac{c+1}{2}\delta \tilde{S}^x_{{\rm II}3}-\frac{c-1}{2}\delta \tilde{S}^y_{{\rm II}3}$ & $-\frac{c-1}{2}\delta \tilde{S}^x_{{\rm IV}2}-\frac{c+1}{2}\delta \tilde{S}^y_{{\rm IV}2}$ & $-\frac{c-1}{2}\delta \tilde{S}^x_{{\rm II}2}-\frac{c+1}{2}\delta \tilde{S}^y_{{\rm II}2}$ \\
III 2   & 2 & $-\frac{c+1}{2}\delta \tilde{S}^x_{{\rm III}1}-\frac{c-1}{2}\delta \tilde{S}^y_{{\rm III}1}$ & $-\frac{c-1}{2}\delta \tilde{S}^x_{{\rm III}3}-\frac{c+1}{2}\delta \tilde{S}^y_{{\rm III}3}$ & $-\frac{c+1}{2}\delta \tilde{S}^x_{{\rm IV}1}-\frac{c-1}{2}\delta \tilde{S}^y_{{\rm IV}1}$ & $\frac{c-1}{2}\delta \tilde{S}^x_{{\rm II}3}+\frac{c+1}{2}\delta \tilde{S}^y_{{\rm II}3}$ & $-s\delta \tilde{S}^x_{{\rm IV}2}-s\delta \tilde{S}^y_{{\rm IV}2}$ & $s\delta \tilde{S}^x_{{\rm II}2}+s\delta \tilde{S}^y_{{\rm II}2}$ \\
  & 3 & $\frac{c-1}{2}\delta \tilde{S}^x_{{\rm III}1}+\frac{c+1}{2}\delta \tilde{S}^y_{{\rm III}1}$ & $s\delta \tilde{S}^x_{{\rm III}3}+s\delta \tilde{S}^y_{{\rm III}3}$ & $-\frac{c-1}{2}\delta \tilde{S}^x_{{\rm IV}1}-\frac{c+1}{2}\delta \tilde{S}^y_{{\rm IV}1}$ & $-s\delta \tilde{S}^x_{{\rm II}3}-s\delta \tilde{S}^y_{{\rm II}3}$ & $-\frac{c+1}{2}\delta \tilde{S}^x_{{\rm IV}2}-\frac{c-1}{2}\delta \tilde{S}^y_{{\rm IV}2}$ & $-\frac{c+1}{2}\delta \tilde{S}^x_{{\rm II}2}-\frac{c-1}{2}\delta \tilde{S}^y_{{\rm II}2}$ \\
 & & & & & & & \\
 \hline
 & & & & & & & \\
   & 1 & $-s\delta \tilde{S}^x_{{\rm IV}1}-s\delta \tilde{S}^y_{{\rm IV}1}$ & $-\frac{c+1}{2}\delta \tilde{S}^x_{{\rm IV}3}-\frac{c-1}{2}\delta \tilde{S}^y_{{\rm IV}3}$ & $-s \delta \tilde{S}^x_{{\rm III}1}-s\delta \tilde{S}^y_{{\rm III}1}$ & $-\frac{c+1}{2}\delta \tilde{S}^x_{{\rm I}3}-\frac{c-1}{2}\delta \tilde{S}^y_{{\rm I}3}$ & $\frac{c-1}{2}\delta \tilde{S}^x_{{\rm III}2}+\frac{c+1}{2}\delta \tilde{S}^y_{{\rm III}2}$ & $\frac{c-1}{2}\delta \tilde{S}^x_{{\rm I}2}+\frac{c+1}{2}\delta \tilde{S}^y_{{\rm I}2}$ \\
IV 2   & 2 & $-\frac{c+1}{2}\delta \tilde{S}^x_{{\rm IV}1}-\frac{c-1}{2}\delta \tilde{S}^y_{{\rm IV}1}$ & $-\frac{c-1}{2}\delta \tilde{S}^x_{{\rm IV}3}-\frac{c+1}{2}\delta \tilde{S}^y_{{\rm IV}3}$ & $-\frac{c+1}{2}\delta \tilde{S}^x_{{\rm III}1}-\frac{c-1}{2}\delta \tilde{S}^y_{{\rm III}1}$ & $\frac{c-1}{2}\delta \tilde{S}^x_{{\rm I}3}+\frac{c+1}{2}\delta \tilde{S}^y_{{\rm I}3}$ & $s\delta \tilde{S}^x_{{\rm III}2}+s\delta \tilde{S}^y_{{\rm III}2}$ & $-s\delta \tilde{S}^x_{{\rm I}2}-s\delta \tilde{S}^y_{{\rm I}2}$ \\
  & 3 & $-\frac{c-1}{2}\delta \tilde{S}^x_{{\rm IV}1}-\frac{c+1}{2}\delta \tilde{S}^y_{{\rm IV}1}$ & $-s\delta \tilde{S}^x_{{\rm IV}3}-s\delta \tilde{S}^y_{{\rm IV}3}$ & $\frac{c-1}{2}\delta \tilde{S}^x_{{\rm III}1}+\frac{c+1}{2}\delta \tilde{S}^y_{{\rm III}1}$ & $s\delta \tilde{S}^x_{{\rm I}3}+s\delta \tilde{S}^y_{{\rm I}3}$ & $-\frac{c+1}{2}\delta \tilde{S}^x_{{\rm III}2}-\frac{c-1}{2}\delta \tilde{S}^y_{{\rm III}2}$ & $-\frac{c+1}{2}\delta \tilde{S}^x_{{\rm I}2}-\frac{c-1}{2}\delta \tilde{S}^y_{{\rm I}2}$ \\
 & & & & & & & \\
 \hline\hline
 & & & & & & & \\
   & 1 & $s\delta \tilde{S}^x_{{\rm I}1}+s\delta \tilde{S}^y_{{\rm I}1}$ & $\frac{c-1}{2}\delta \tilde{S}^x_{{\rm I}2}+\frac{c+1}{2}\delta \tilde{S}^y_{{\rm I}2}$ & $-s\delta \tilde{S}^x_{{\rm III}1}-s\delta \tilde{S}^y_{{\rm III}1}$ & $-\frac{c-1}{2}\delta \tilde{S}^x_{{\rm IV}2}-\frac{c+1}{2}\delta \tilde{S}^y_{{\rm IV}2}$ & $-\frac{c+1}{2}\delta \tilde{S}^x_{{\rm III}3}-\frac{c-1}{2}\delta \tilde{S}^y_{{\rm III}3}$ & $-\frac{c+1}{2}\delta \tilde{S}^x_{{\rm IV}3}-\frac{c-1}{2}\delta \tilde{S}^y_{{\rm IV}3}$ \\
I 3  & 2 & $-\frac{c+1}{2}\delta \tilde{S}^x_{{\rm I}1}-\frac{c-1}{2}\delta \tilde{S}^y_{{\rm I}1}$ & $-s\delta \tilde{S}^x_{{\rm I}2}-s\delta \tilde{S}^y_{{\rm I}2}$ & $-\frac{c+1}{2}\delta \tilde{S}^x_{{\rm III}1}-\frac{c-1}{2}\delta \tilde{S}^y_{{\rm III}1}$ & $-s\delta \tilde{S}^x_{{\rm IV}2}-s\delta \tilde{S}^y_{{\rm IV}2}$ & $-\frac{c-1}{2}\delta \tilde{S}^x_{{\rm III}3}-\frac{c+1}{2}\delta \tilde{S}^y_{{\rm III}3}$ & $-\frac{c-1}{2}\delta \tilde{S}^x_{{\rm IV}3}-\frac{c+1}{2}\delta \tilde{S}^y_{{\rm IV}3}$ \\
  & 3 & $-\frac{c-1}{2}\delta \tilde{S}^x_{{\rm I}1}-\frac{c+1}{2}\delta \tilde{S}^y_{{\rm I}1}$ & $-\frac{c+1}{2}\delta \tilde{S}^x_{{\rm I}2}-\frac{c-1}{2}\delta \tilde{S}^y_{{\rm I}2}$ & $\frac{c-1}{2}\delta \tilde{S}^x_{{\rm III}1}+\frac{c+1}{2}\delta \tilde{S}^y_{{\rm III}1}$ & $-\frac{c+1}{2}\delta \tilde{S}^x_{{\rm IV}2}-\frac{c-1}{2}\delta \tilde{S}^y_{{\rm IV}2}$ & $s\delta \tilde{S}^x_{{\rm III}3}+s\delta \tilde{S}^y_{{\rm III}3}$ & $-s\delta \tilde{S}^x_{{\rm IV}3}-s\delta \tilde{S}^y_{{\rm IV}3}$ \\
 & & & & & & & \\
  \hline
 & & & & & & & \\
    & 1 & $s\delta \tilde{S}^x_{{\rm II}1}+s\delta \tilde{S}^y_{{\rm II}1}$ & $-\frac{c-1}{2}\delta \tilde{S}^x_{{\rm II}2}-\frac{c+1}{2}\delta \tilde{S}^y_{{\rm II}2}$ & $-s\delta \tilde{S}^x_{{\rm IV}1}-s\delta \tilde{S}^y_{{\rm IV}1}$ & $\frac{c-1}{2}\delta \tilde{S}^x_{{\rm III}2}+\frac{c+1}{2}\delta \tilde{S}^y_{{\rm III}2}$ & $-\frac{c+1}{2}\delta \tilde{S}^x_{{\rm IV}3}-\frac{c-1}{2}\delta \tilde{S}^y_{{\rm IV}3}$ & $-\frac{c+1}{2}\delta \tilde{S}^x_{{\rm III}3}-\frac{c-1}{2}\delta \tilde{S}^y_{{\rm III}3}$ \\
II 3  & 2 & $-\frac{c+1}{2}\delta \tilde{S}^x_{{\rm II}1}-\frac{c-1}{2}\delta \tilde{S}^y_{{\rm II}1}$ & $s\delta \tilde{S}^x_{{\rm II}2}+s\delta \tilde{S}^y_{{\rm II}2}$ & $-\frac{c+1}{2}\delta \tilde{S}^x_{{\rm IV}1}-\frac{c-1}{2}\delta \tilde{S}^y_{{\rm IV}1}$ & $s\delta \tilde{S}^x_{{\rm III}2}+s\delta \tilde{S}^y_{{\rm III}2}$ & $-\frac{c-1}{2}\delta \tilde{S}^x_{{\rm IV}3}-\frac{c+1}{2}\delta \tilde{S}^y_{{\rm IV}3}$ & $-\frac{c-1}{2}\delta \tilde{S}^x_{{\rm III}3}-\frac{c+1}{2}\delta \tilde{S}^y_{{\rm III}3}$ \\
  & 3 & $\frac{c-1}{2}\delta \tilde{S}^x_{{\rm II}1}+\frac{c+1}{2}\delta \tilde{S}^y_{{\rm II}1}$ & $-\frac{c+1}{2}\delta \tilde{S}^x_{{\rm II}2}-\frac{c-1}{2}\delta \tilde{S}^y_{{\rm II}2}$ & $-\frac{c-1}{2}\delta \tilde{S}^x_{{\rm IV}1}-\frac{c+1}{2}\delta \tilde{S}^y_{{\rm IV}1}$ & $-\frac{c+1}{2}\delta \tilde{S}^x_{{\rm III}2}-\frac{c-1}{2}\delta \tilde{S}^y_{{\rm III}2}$ & $-s\delta \tilde{S}^x_{{\rm IV}3}-s\delta \tilde{S}^y_{{\rm IV}3}$ & $s\delta \tilde{S}^x_{{\rm III}3}+s\delta \tilde{S}^y_{{\rm III}3}$ \\
 & & & & & & & \\
  \hline
 & & & & & & & \\
    & 1 & $-s\delta \tilde{S}^x_{{\rm III}1}-s\delta \tilde{S}^y_{{\rm III}1}$ & $\frac{c-1}{2}\delta \tilde{S}^x_{{\rm III}2}+\frac{c+1}{2}\delta \tilde{S}^y_{{\rm III}2}$ & $s\delta \tilde{S}^x_{{\rm I}1}+s\delta \tilde{S}^y_{{\rm I}1}$ & $-\frac{c-1}{2}\delta \tilde{S}^x_{{\rm II}2}-\frac{c+1}{2}\delta \tilde{S}^y_{{\rm II}2}$ & $-\frac{c+1}{2}\delta \tilde{S}^x_{{\rm I}3}-\frac{c-1}{2}\delta \tilde{S}^y_{{\rm I}3}$ & $-\frac{c+1}{2}\delta \tilde{S}^x_{{\rm II}3}-\frac{c-1}{2}\delta \tilde{S}^y_{{\rm II}3}$ \\
III 3  & 2 & $-\frac{c+1}{2}\delta \tilde{S}^x_{{\rm III}1}-\frac{c-1}{2}\delta \tilde{S}^y_{{\rm III}1}$ & $s\delta \tilde{S}^x_{{\rm III}2}+s\delta \tilde{S}^y_{{\rm III}2}$ & $-\frac{c+1}{2}\delta \tilde{S}^x_{{\rm I}1}-\frac{c-1}{2}\delta \tilde{S}^y_{{\rm I}1}$ & $s\delta \tilde{S}^x_{{\rm II}2}+s\delta \tilde{S}^y_{{\rm II}2}$ & $\frac{c-1}{2}\delta \tilde{S}^x_{{\rm I}3}+\frac{c+1}{2}\delta \tilde{S}^y_{{\rm I}3}$ & $\frac{c-1}{2}\delta \tilde{S}^x_{{\rm II}3}+\frac{c+1}{2}\delta \tilde{S}^y_{{\rm II}3}$ \\
  & 3 & $\frac{c-1}{2}\delta \tilde{S}^x_{{\rm III}1}+\frac{c+1}{2}\delta \tilde{S}^y_{{\rm III}1}$ & $-\frac{c+1}{2}\delta \tilde{S}^x_{{\rm III}2}-\frac{c-1}{2}\delta \tilde{S}^y_{{\rm III}2}$ & $-\frac{c-1}{2}\delta \tilde{S}^x_{{\rm I}1}-\frac{c+1}{2}\delta \tilde{S}^y_{{\rm I}1}$ & $-\frac{c+1}{2}\delta \tilde{S}^x_{{\rm II}2}-\frac{c-1}{2}\delta \tilde{S}^y_{{\rm II}2}$ & $s\delta \tilde{S}^x_{{\rm I}3}+s\delta \tilde{S}^y_{{\rm I}3}$ & $-s\delta \tilde{S}^x_{{\rm II}3}-s\delta \tilde{S}^y_{{\rm II}3}$ \\
 & & & & & & & \\
  \hline
 & & & & & & & \\
    & 1 & $-s\delta \tilde{S}^x_{{\rm IV}1}-s\delta \tilde{S}^y_{{\rm IV}1}$ & $-\frac{c-1}{2}\delta \tilde{S}^x_{{\rm IV}2}-\frac{c+1}{2}\delta \tilde{S}^y_{{\rm IV}2}$ & $s\delta \tilde{S}^x_{{\rm II}1}+s\delta \tilde{S}^y_{{\rm II}1}$ & $\frac{c-1}{2}\delta \tilde{S}^x_{{\rm I}2}+\frac{c+1}{2}\delta \tilde{S}^y_{{\rm I}2}$ & $-\frac{c+1}{2}\delta \tilde{S}^x_{{\rm II}3}-\frac{c-1}{2}\delta \tilde{S}^y_{{\rm II}3}$ & $-\frac{c+1}{2}\delta \tilde{S}^x_{{\rm I}3}-\frac{c-1}{2}\delta \tilde{S}^y_{{\rm I}3}$ \\  
IV 3  & 2 & $-\frac{c+1}{2}\delta \tilde{S}^x_{{\rm IV}1}-\frac{c-1}{2}\delta \tilde{S}^y_{{\rm IV}1}$ & $-s\delta \tilde{S}^x_{{\rm IV}2}-s\delta \tilde{S}^y_{{\rm IV}2}$ & $-\frac{c+1}{2}\delta \tilde{S}^x_{{\rm II}1}-\frac{c-1}{2}\delta \tilde{S}^y_{{\rm II}1}$ & $-s\delta \tilde{S}^x_{{\rm I}2}-s\delta \tilde{S}^y_{{\rm I}2}$ & $\frac{c-1}{2}\delta \tilde{S}^x_{{\rm II}3}+\frac{c+1}{2}\delta \tilde{S}^y_{{\rm II}3}$ & $\frac{c-1}{2}\delta \tilde{S}^x_{{\rm I}3}+\frac{c+1}{2}\delta \tilde{S}^y_{{\rm I}3}$ \\
  & 3 & $-\frac{c-1}{2}\delta \tilde{S}^x_{{\rm IV}1}-\frac{c+1}{2}\delta \tilde{S}^y_{{\rm IV}1}$ & $-\frac{c+1}{2}\delta \tilde{S}^x_{{\rm IV}2}-\frac{c-1}{2}\delta \tilde{S}^y_{{\rm IV}2}$ & $\frac{c-1}{2}\delta \tilde{S}^x_{{\rm II}1}+\frac{c+1}{2}\delta \tilde{S}^y_{{\rm II}1}$ & $-\frac{c+1}{2}\delta \tilde{S}^x_{{\rm I}2}-\frac{c-1}{2}\delta \tilde{S}^y_{{\rm I}2}$ & $-s\delta \tilde{S}^x_{{\rm II}3}-s\delta \tilde{S}^y_{{\rm II}3}$ & $s\delta \tilde{S}^x_{{\rm I}3}+s\delta \tilde{S}^y_{{\rm I}3}$ \\
 & & & & & & & \\
  \hline
\end{tabular}
}
\end{table}

\end{document}